%% file: ex_article.tex
\documentclass[onefignum,onetabnum]{siamart190516}

\input{ex_shared}

\ifpdf
\hypersetup{
  pdftitle={High-Dimensional Dynamic Stochastic Model Representation},
  pdfauthor={Aryan Eftekhari, Simon Scheidegger}
}
\fi

\usepackage{multirow}
\usepackage{amsmath}
\usepackage{pgf}
\usepackage{xcolor}
\usepackage{mathtools}
\usepackage{braket}
\usepackage{threeparttable}
\usepackage{comment}
\usepackage{hyperref}
\hypersetup{%
    colorlinks=true,       
    linkcolor=red,          
    citecolor=blue,        
    filecolor=magenta,      
    urlcolor=cyan           
}
\usepackage{tikz}
\usetikzlibrary{arrows}
\numberwithin{equation}{section}
\usetikzlibrary{positioning}
\usepackage{calrsfs}
\DeclareMathAlphabet{\pazocal}{OMS}{zplm}{m}{n}
\usepackage{amssymb}
\usepackage{algorithm}
\usepackage{algorithmic}
\usepackage{bigints}

\newcommand{\spn}[1]{\text{span}\left\{ #1 \right\} }

\newcommand{\eval}[2]{|_{#1=#2}}
\newcommand\norm[2]{\left\lVert #1 \right\rVert_{#2}}

\newcommand{\ignore}[1]{}
\newcommand{\card}[1]{|#1|}
\newcommand{\ekk}[0]{\mathcal{K}}

\newcommand{\argmin}[1]{\underset{#1}{\text{argmin}} }

\newcommand{\bb}[1]{\boldsymbol{\mathbf{#1}}}

\newcommand{\bbb}[1]{\bar{\boldsymbol{\mathrm{#1}}}}

\DeclareMathAlphabet{\mathcal}{OMS}{cmsy}{m}{n}

\begin{document}

\maketitle

\begin{abstract}
We propose a scalable method for computing global solutions of nonlinear, high-dimensional dynamic stochastic economic models.
First, within a time iteration framework, we approximate economic policy functions using an adaptive, high-dimensional model representation scheme, combined with adaptive sparse grids to address the ubiquitous challenge of the curse of dimensionality. 
Moreover, the adaptivity within the individual component functions increases sparsity since grid points are added only where they are most needed, that is, in regions with steep gradients or at nondifferentiabilities. 
Second, we introduce a performant vectorization scheme for the interpolation compute kernel. 
Third, the algorithm is hybrid parallelized, leveraging both distributed- and shared-memory architectures.  
We observe significant speedups over the state-of-the-art techniques, and almost ideal strong scaling up to at least $1,000$ compute nodes of a Cray XC$50$ system at the Swiss National Supercomputing Center.
Finally, to demonstrate our method's broad applicability, we compute global solutions to two variates of a high-dimensional international real business cycle model up to $300$ continuous state variables.
In addition, we highlight a complementary advantage of the framework, which allows for a priori analysis of the model complexity.
\end{abstract}



\section{Introduction}\label{sec:1}
Motivated by empirical observations, features such as heterogeneity, interconnectedness, and uncertainty have become vital ingredients for capturing the salient features in contemporary dynamic economic models. 
In macroeconomics for instance, heterogeneity between types of agents, such as hand-to-mouth and non-hand-to-mouth consumers (see, e.g.,~\cite{kaplanetal_2018}), financial frictions such as collateral constraints (see, e.g.,~\cite{kublerschmedders_2003}), or distributional channels (see, e.g.,~\cite{kruegeretal_2016}) have become widely recognized as essential components for modern macroeconomic models.
The need for heterogeneity has recently become even more evident during the current \textit{COVID-19} pandemic, where unprecedented policy actions have to be taken to mitigate this once-in-a-century event (see,e.g.,~\cite{Hsiang2020}).

To address today's key questions in economics quantitatively, one quickly ends up with an intricate formal structure that relies on considering so-called \textit{recursive equilibria}~\cite{stokey1989recursive,ljungqvist2004recursive}.
In such equilibria, a potentially high-dimensional \textit{state variable} $\bb{x} \in X  \subset \mathbb{R}^d$ represents the state of the economy, $d$ is the dimensionality of the state space, and a time-invariant \emph{optimal policy function} $p : X \rightarrow Y \subset \mathbb{R}^m$, the desired unknown, captures the model dynamics and can be characterized as the solution to a functional equation that reads~\cite{FERNANDEZVILLAVERDE2016527}:
\begin{equation}
    \mathcal{H}(p)=\mathbf{0}.
\label{eq:DynamicEq}
\end{equation}
This abstract description nests various characterizations of recursive equilibria, and in particular, the widespread case where the operator $\mathcal{H}$ captures discrete-time first-order equilibrium conditions, the focal point of this paper.

A standard method for solving such \textit{dynamic stochastic economic models} is the so-called \textit{time iteration} algorithm~\cite{coleman1990}, which computes the recursive equilibrium of a dynamic economic model by guessing a policy function and iteratively updating it using the first-order equilibrium conditions of the model.
However, solving for global solutions\footnote{A \emph{global solution} adheres to the model equilibrium conditions throughout the entire state space---that is, the computational domain, whereas a \emph{local solution} is only concerned with the local approximation around a steady state.} to models with substantial heterogeneity and highly nonlinear policy functions is very costly for two key reasons. First, no matter which model characterization is used, the \textit{curse of dimensionality}~\cite{Bellman1961AdaptiveTour} imposes a roadblock as soon as $X$ is of a higher dimension (say $d>3$), and a global solution has to be computed, where the equilibrium conditions need to be satisfied throughout the entire computational domain. A grid-based solution technique that relies on a \textit{naive} tensor-product construction will require $\mathcal{O}(M^d)$ points when $M$ points are needed in each dimension.
This exponential growth makes tensor-product grids infeasible as soon as $M$ and $d$ reach moderate levels. Second, at each grid point, a system of nonlinear equations must be solved. When solving the system of equations at a given grid point, one needs to frequently interpolate from the function computed in the previous iteration step. These interpolation operations can account for up to $99\%$ of the overall compute time for solving the system of equations~\cite{IPDPS2018}. These two impediments make it difficult to achieve an acceptable time-to-solution which can quickly reach the order of days on modern supercomputing facilities. As a result, present methods frequently fall well short of including as much heterogeneity as a reasonable modeling choice would imply.\footnote{For example,~\cite{Krueger20041411} examined the welfare implications of social security reforms using a model in which one period amounted to six years rather than one, thus lowering the number of adult cohorts and therefore the problem's dimensionality by a factor of six. Similarly, international real business cycle models often incorporate only a small number of countries. For instance,~\cite{bengui2013} studied cross-country risk-sharing at the business cycle frequency using a two-country model with one focus country and the rest of the world. By reducing the problem's dimensionality in this way, valuable qualitative insights can be gained. However, in order to obtain reliable quantitative results or simply to assess the robustness of qualitative findings, it is frequently necessary to consider problems of larger dimensions.}

To deal with the ever-increasing complexity of state-of-the-art dynamic stochastic economic models, we propose in this work a generic, scalable, and flexible computational framework which can efficiently address the previously noted bottlenecks. 
Building on~\cite{IPDPS2018,brummscheidegger_2017,Brumm201512} and~\cite{Eftekhari:2017:PDD:3093172.3093234}, we specifically contribute (i) an adaptive high-dimensional model representation scheme that is coupled with an adaptive sparse grid algorithm applicable for recursively formulated economic models that significantly reduces the number of grid points in the approximation and the time needed for each function evaluation, (ii) adaptivity criteria which can be used as an on-the-fly analysis tool elucidating the complexity of the model under consideration, (iii) a vectorized implementation for performant interpolation, and (iv) a hybrid parallelized time iteration solution framework fit for virtually any dynamic stochastic economic model. 
Finally (v), we deploy our solution framework at the Swiss National Supercomputing Center (CSCS) to solve highly nonlinear dynamic stochastic economic models of up to $300$ dimensions globally.

The grid point reduction is achieved by combining adaptive sparse grids (adaptive SGs; see, e.g.,~\cite{Bungartz2003MultivariateGrids,Pfluger2010SpatiallyProblems,Ma:2009:AHS:1514432.1514547}) with a dimensional decomposition (DD) framework that is based on high-dimensional model representation (see, e.g.,~\cite{genyuan,Ma2010AnEquations,Yang2012AdaptiveFlows}). 
Finally, the time-to-solution is substantially accelerated by using a hybrid parallelization scheme (i.e., using both distributed- and shared-memory hardware) combined with a novel vectorization approach for fast interpolation on the policy functions.

SGs can alleviate the curse of dimensionality to some extent, allowing one to tackle models that incorporate rich economic settings, including international real business cycle (IRBC) models of up to $50$ countries, that is, $100$ dimensions~\cite{brummscheidegger_2017}. Furthermore, adaptive SGs can resolve steep gradients or nondifferentiabilities efficiently, making them useful in the context of solving a broad range of mid-scale economic models (say $d<20$) with non-smooth policy functions that arise, for example, in the presence of collateral constraints on borrowing (see~\cite{SG_in_econ_handbook} for a review on the use of adaptive SGs in economics and finance).

However, the limitations of adaptive SGs become evident when one considers highly nonlinear economic models of much more heterogeneity, for example, IRBC models that consist of dozens of countries that face irreversible investment constraints. 
In such situations, adaptive SGs are no longer an applicable solution method, as the number of grid points increases substantially with the approximation quality or, equivalently, with the resolution of the grid. Furthermore, in high-dimensional SGs, access times of the data structures become computationally expensive (see, e.g.,~\cite{ppopp027s-murarasu}).

The noted issues can be surpassed by combining DD (see, e.g.,~\cite{Li2012GeneralVariables,Rabitz1999GeneralRepresentations,Hooker}) with adaptive SGs, which we refer to as DDSG. 
The core idea of DD is to approximate a function by decomposing it into a series of lower-dimensional functions. This decomposition gives a way to represent, for example, in the simplest case, a $300$-dimensional function as a summation $300$ one-dimensional functions. We will focus on one variate of DD referred to in the literature as High-Dimensional Model Representation (HDMR). 
The HDMR technique has been applied to multiple fields including chemistry (see, e.g.,~\cite{Yang2012AdaptiveFlows}), physics (see, e.g.,~\cite{Ma2010AnEquations}), and machine-learning (see, e.g.,~\cite{DBLP:journals/corr/abs-1906-02005}). 
With this said, and to the best of our knowledge, HDMR so far seems to have gone unrecognized in the field of economics until now.

The remainder of this paper is organized as follows. In Section~\ref{sec:lit}, we provide a very brief review of the related literature on global solution techniques for high-dimensional dynamic economic models. In Section~\ref{sec:2}, we describe the abstract structure of the models we aim to solve with our proposed method and specify a conceptually simple yet computationally demanding economic test case---the IRBC model. 
The latter has become the de-facto workhorse for studying methods for solving high-dimensional economic models, as its dimensionality can be scaled up in a straightforward and meaningful way, as it just depends linearly on the number of
countries considered. In Section~\ref{sec:3}, we detail the proposed DDSG method. Next, in Section~\ref{sec:4}, we embed DDSG in a time iteration algorithm and discuss the hybrid parallelization scheme of the complete method. In Section~\ref{sec:5}, we support the performance claims and applicability of the proposed solution framework by providing a series of unit tests and global solution results in two different configurations of high-dimensional IRBC models. Finally, in Section~\ref{sec:6} we conclude.

\section{A brief literature review on global solution methods}
\label{sec:lit}
Over the past two decades, there have been significant advancements in the development of algorithms and numerical tools to compute global solutions for high-dimensional dynamic stochastic economic models (see~\cite{MALIAR2014325,FERNANDEZVILLAVERDE2016527} for recent reviews). To overcome the inherent curse of dimensionality, the computational economics community has pursued two main strands of research: i) SG-based solution algorithms and ii) grid-free methods. 
SG methods~\cite{Bungartz.Griebel:2004} are a mathematically well-studied, systematic way to tackle the numerical difficulties that arise in dynamic economic models due to the high-dimensional state spaces. However, they typically fail in real applications if the dimensionality of a highly nonlinear model exceeds about $20$ (see~\cite{SG_in_econ_handbook} for a recent review). However, such problem sizes are often required from a theoretical point of view, for instance, in annually calibrated multi-country overlapping generation models with borrowing constraints---a $240$-dimensional problem in its minimal formulation. In contrast, DDSG can, as we show below, handle problem sizes of at least $300$ continuous state variables.
Grid-free approaches have been proposed, for example, by~\cite{den1990solving} and have lately become more powerful by leveraging the rapid developments in machine learning. In~\cite{rennerscheidegger_2020,kotlikoffetal_2020_WP} and~\cite{scheideggerbilinois_2019,kublerscheidegger_2018_WP}, the authors combine Gaussian processes with reinforcement learning and active subspaces, respectively. 
However, the combination of these methods typically cannot deal in a straightforward manner with frictions such as irreversible investments and the related nonlinearities (as presented in the IRBC model in Section~\ref{sec:FOC_nonsmooth}), whereas DDSG can. Other research in computational economics has recently applied deep neural networks to compute global solutions to high-dimensional dynamic stochastic models (see, e.g.,~\cite{azinovic_et_al_2019,MALIAR202176,fernandez-villaverdehurtadonuno_2019_WP}). However, while these methods could be considered an alternative to the work presented here, they nowadays still suffer from several drawbacks that limit their general applicability. Their convergence properties, for example, with respect to the network architecture, are still poorly understood, thus often requiring a substantial amount of hyper-parameter tuning for a successful model solution. In contrast, DDSGs provide a transparent way to handle high-dimensional models. Finally, nesting the DD approach with the time iteration algorithm is not restricted to SGs. Thus, alternative solution methods for low-dimensional economic models could also be scaled up in a relatively straightforward way if embedded in DD, in turn providing a simple means to extend the boundaries for research.

\section{Large-scale dynamic stochastic economic models} \label{sec:2}
This section outlines the types of models and the recursive solution techniques that we use below to demonstrate the versatility, accuracy, and computational scalability of the method introduced in this paper. To do so, we proceed in two main steps. First, we begin in Section~\ref{sec:2.1} by briefly describing~\textit{smooth} and~\textit{non-smooth} variants of the IRBC model as concrete test cases\footnote{As custom in the literature, we denote an IRBC model with no kinks in the policies as \textit{smooth}, whereas we name it \textit{non-smooth} if there exist non-differentiabilities in the latter functions.}, thereby closely following~\cite{Brumm201512}. The latter has recently become a workhorse model for studying methods for solving high-dimensional dynamic stochastic models (see, e.g.,~\cite{Juillard2011178,RePEc:eee:dyncon:v:35:y:2011:i:2:p:175-177}, and references therein). The IRBC model is straightforward to explain, has a unique solution~\cite{stokey1989recursive}, and its dimensionality can be scaled up in a meaningful way as it just depends linearly on the number of countries considered. This property of the model allows us to concentrate on the computational issues of handling high-dimensional state spaces. To demonstrate that we can also handle non-smooth problems, we also consider a version of the IRBC model where investment is irreversible. Second, we present in Section~\ref{sec:2.2}---based on the example of the IRBC models---how a recursive equilibrium can be computed by applying the time iteration algorithm and discuss the computational challenges associated with solving for global solutions of large-scale dynamic stochastic economic models, and which motivate the development of our proposed DDSG method.

\subsection{A scalable test case: the international real business cycle model} \label{sec:2.1}
In the IRBC model we are considering as test case, time $t$ is discrete, there are $N$ countries, $j=1,\dots,N$, each using its accumulated capital stock, $k_{j,t}\in \mathbb{R}_{+}$, to produce an output good, which can be used for investment, $\mathrm{I}_{j,t}$, and for consumption, $c_{j,t}\in \mathbb{R}_{+}$, generating utility, $\mathrm{u}_{j}(c_{j,t})$.
In the model formulation we follow, complete markets are assumed~\cite{Kollmann2011186}. Thus, a social planner solves the following (infinite-horizon) optimization problem:
\begin{align} \label{eq:social_planner}
\max_{ \{{c}_{j,t},{k}_{j,t+1}\} } & \left\{ \mathbb{E}_0 \left[ \sum_{j=1}^N {\tau}_{j} \sum_{t=0}^{\infty} \beta^t \mathrm{u}_j({c}_{j,t}) \right] \right\} \nonumber \\
\text{subject to:}\;
& (1)\;\sum_{j=1}^{N} \mathrm{R}({a}_{j,t},{k}_{j,t},{k}_{j,t+1},{c}_{j,t}) \geqslant 0 ,\; \forall \, t \nonumber
\\
& (2)\;\mathrm{I}_{j,t}({k}_{j,t},{k}_{j,t+1}) \geqslant 0 ,\; \forall \, \{t , j \},
\end{align}
where $a_{j,t}$ denotes productivity, $\tau_{j}$ are the welfare weights, $\beta$ is the discount factor, $\mathbb{E}_0$ is the expectation conditional on the information available at $t=0$, and where the initial capital stocks ${\mathbf{k}}_0 \in \mathbb{R}_{+}^N$ and productivity levels ${\mathbf{a}}_0 \in \mathbb{R}_{+}^N$ are assumed to be given. The parameterization for both the smooth and non-smooth IRBC model follows~\cite{Juillard2011178,brummscheidegger_2017} and is reported in Table~\ref{tab:1}.
The first constraint (1) is the so-called \emph{aggregate resource constraint}, while the second constraint (2) enforces \emph{irreversible investments}.\footnote{Investment is irreversible in the sense that it cannot be consumed or used for production in another country---an assumption that is more realistic than perfect reversibility, which is usually assumed to keep the model tractable.} 
We refer to the \emph{smooth} IRBC model where only constraint (1) is used. In contrast, the \emph{non-smooth} IRBC model requires the solution of~\eqref{eq:social_planner} with both constraints (1) and (2). Furthermore, we assume an additive separable per-period utility function that is given by $\mathrm{u}_j(c_{j,t}) =  {c}_{j,t}^{1-\frac{1}{{\gamma}_j}}/({1-\frac{1}{{\gamma}_j}})$. 
The aggregate resource constraint is a function of the production function $\mathrm{Y}$, investment $\mathrm{I}_{j,t}$, the capital adjustment costs $\mathrm{\Gamma}$, and consumption ${c}_{j,t}$:
\begin{align}
 \mathrm{R}({a}_{j,t},{k}_{j,t},{k}_{j,t+1},{c}_{j,t}) &=
 \mathrm{Y}({a}_{j,t},{k}_{j,t})-
\Gamma({k}_{j,t},{k}_{j,t+1})  \!-\! \mathrm{I}({k}_{j,t},{k}_{j,t+1}) - {c}_{j,t},  \nonumber \\
\mathrm{Y}({a}_{j,t},{k}_{j,t}) &= A\cdot {a}_{j,t}\cdot {k}^\alpha_{j,t}, \nonumber \\
\mathrm{I}_{j,t}({k}_{j,t},{k}_{j,t+1} ) &= {k}_{j,t+1} - (1-\delta)\cdot {k}_{j,t}, \, \quad \text{and} \label{eqn:investment} \\
\Gamma({k}_{j,t},{k}_{j,t+1}) &= \frac{\phi}{2}\cdot {k}_{j,t}\cdot \left( \frac{{k}_{j,t+1}}{ {k}_{j,t}} - 1 \right)^2. 
\end{align}
The law of motion of productivity is the sole source of stochasticity in the model and is given by:
\begin{align}\label{eqn:law_of_motion}
\ln {a}_{j,t} = \rho \cdot \ln {a}_{j,t-1} + \sigma \cdot ( {e}_{j,t} +  {e}_t),
\end{align}
where ${e}_{j,t}\sim \mathcal{N}(0,1)$ and ${e}_{t} \sim \mathcal{N}(0,1)$ denote the country-specific, and the global shocks, respectively. Both are assumed to be independent from each other and across time. Thus far, we have considered an infinite horizon problem. However, as indicated previously, economics frequently focuses on recursive equilibria~\cite{stokey1989recursive,ljungqvist2004recursive}, in which the state of the economy is represented by a state variable and the economy's dynamics are given by a time-invariant function of this state (cf.~\eqref{eq:DynamicEq}).
We now briefly describe the recursive structure, that is, the two IRBC model formulations' first-order optimality conditions (FOCs). We direct the reader to~\cite{brummscheidegger_2017} for the derivations.
\begin{table}[t]%
\centering
\caption{Parameterization of the smooth and non-smooth IRBC model.}
\scalebox{0.85}{
\begin{tabular}{l|c|c}
$\textbf{Parameter}$ & $\textbf{Symbol}$ & $\textbf{Value}$ \\ \hline 
Discount factor & $\beta$ & $0.99$ \\ 
Elasticity of intertemporal substitution of country $j$ & $\gamma_j$ & $a+0.75(j-1)/(N-1)$ \\ 
Capital share  &  $\alpha$ &  $0.36$ \\ 
Depreciation &  $\delta$ &  $0.01$ \\ 
Std.\ of $\log$-productivity shocks  &  $\sigma$ &  $0.01$ \\ 
Autocorrelation of $\log$-productivity  &  $\rho$ &  $0.95 $\\ 
Intensity of capital adjustment costs    &  $\phi$  &$ 0.50$  \\ 
Aggregate productivity &  $A$ & $\left( 1 -\beta (1- \delta) \right)/(\alpha \cdot \beta)$  \\ 
Welfare weights  & $\tau_j$  & $A^{1/\gamma_j}$ \\
\end{tabular}
}
\label{tab:1}
\end{table}
%

\subsubsection{Smooth IRBC model - recursive structure} \label{sec:FOC_smooth}
In order to obtain FOCs of the optimization problem stated in equation~\eqref{eq:social_planner}, we need to differentiate the Lagrangian with respect to $c_{j,t}$ and $k_{j,t+1}$. Denoting $\lambda_t$ as the multiplier on the resource constraint at time $t$, and defining the growth rate of capital by $ g_{j,t}=k_{j,t}/k_{j,t-1} -1 $, we obtain a system of $N$ equilibrium conditions that have to hold at each $t$ and for all countries $j$:
\begin{multline}
 \lambda_t  \left( 1 + \phi   g_{j,t+1} \right) - \\
\beta  \mathbb{E}_t \left[ \lambda_{t+1} \left( a_{j,t+1}  A 
  \alpha  (k_{j,t+1})^{\alpha-1} + (1 - \delta)   
+ \frac{\phi}{2} g_{j,t+2}   \left( g_{j,t+2} + 2  \right)  \right)  \right] = 0. 
\label{eq:model1}
\end{multline}
where $\mathbb{E}_t$ is the expectation conditional on the information available at $t$. Furthermore, the aggregate resource constraint (holding with equality due to strictly increasing per-period utility assumed) reads as follows:
\begin{equation}
\sum_{j=1}^{N} \left( a_{j,t}  A  (k_{j,t})^{\alpha}  + 
k_{j,t}  \left( (1-\delta) - \frac{\phi}{2}  (g_{j,t+1})^2 \right) - k_{j,t+1} - 
\left(\frac{\lambda_t}{\tau_j}\right)^{-\gamma_j} \right) = 0 ,
\label{eq:model2}
\end{equation}
where we use the fact that $c_{j,t}= ({\lambda_t}/{\tau_j})^{-\gamma^j}$ holds at an optimal choice~\cite{brummscheidegger_2017}. To explicitly point out the link to the abstract definition of a recursive equilibrium~\eqref{eq:DynamicEq} (and the time iteration Algorithm~\ref{alg:1} described in section~\ref{sec:2.2} below), note that the smooth IRBC model presented here has a $(d=2N)$-dimensional state space. As a reminder, the state variables are given by $ \mathbf{x}_{t} =\left(\bb{a}_{t},\bb{k}_t\right) \in \mathbb{R}^{2N}$, where $\bb{a}_{t}=(a_{1,t},\ldots,a_{N,t})$ and $\bb{k}_{t}=(k_{1,t},\ldots,k_{N,t}) $, are the productivity and capital stock of country $j $. Furthermore, the optimal, time-invariant policy $p:\mathbb{R}^{2N}\rightarrow\mathbb{R}^{N+1}$---the desired unknown---maps the current state into policies as $p(\mathbf{x}_t)=(\bb{k}_{t+1},\lambda_t)$.
Note that the investment choices determine the capital stock of the next period (i.e., $t+1$) in a deterministic way through~\eqref{eqn:investment}. In contrast, the law of motion of productivity,~\eqref{eqn:law_of_motion}, is stochastic.
Taken together,~\eqref{eqn:investment} and~\eqref{eqn:law_of_motion} specify the distribution of $\mathbf{x}_{t+1}$.\footnote{Throughout this paper, we compute the expectations by a simple yet fast monomial quadrature rule (see, e.g., \cite{judd1998numerical}, Sec. 7.5).} 

\subsubsection{Non-smooth IRBC model - recursive structure} \label{sec:FOC_nonsmooth}
To demonstrate the strength of our proposed algorithm to deal with high-dimensional and highly nonlinear models, we consider next a variant of the IRBC model where investment is irreversible, which in turn leads to non-smooth optimal policies. More precisely, we assume that investment cannot be negative (cf. constraint (2) in equation~\eqref{eq:social_planner}). 
As a direct consequence, we have to solve a system of $2N+1$ equilibrium conditions. These conditions now include the Karush--Kuhn--Tucker (KKT) multiplier, $\mu_{j,t}$, for the irreversibility constraint. The optimality conditions for investment in capital as well as the irreversibility assumption for investment in each country $j$, and the associated complementary conditions, read as: 
\begin{multline}
\lambda_t  \left( 1 + \phi  g_{j,t+1} \right)  - \mu_{j,t}  
- \\ \beta \mathbb{E}_t \left[ \lambda_{t+1} \left( a_{j,t+1}  A 
 \alpha  (k_{j,t+1})^{\alpha-1} + 1 - \delta   
+ \frac{\phi}{2} g_{j,t+2}   \left( g_{j,t+2} + 2  \right)  \right)   - (1 - \delta) \mu_{j,t+1} \right] = 0,  \\
0 \leq  \mu_{j,t}  \perp \left( k_{j,t+1} -  k_{j,t} (1-\delta)\right)  \geq 0.  
\label{eq:modelkink1}
\end{multline}
In addition, the aggregate resource constraint holds again. The state variables of this non-smooth IRBC model are again given by $ \mathbf{x}_{t} = (\bb{a}_{t},\bb{k}_t)$.
The optimal, time-invariant policy $p:\mathbb{R}^{2N}\rightarrow\mathbb{R}^{2N+1}$ now maps the current state into policies as $p(\mathbf{x}_t)=(\bb{k}_{t+1},\bb{\mu}_{t},\lambda_{t})$ where $\bb{\mu}_{t}=(\mu_{1,t}, \dots, \mu_{N,t})$.
As in the previous Section~\ref{sec:FOC_smooth}, all the policies will have to be determined by iterating on~\eqref{eq:modelkink1} and the aggregate resource constraint.

\subsection{The time iteration algorithm and its computational challenges} \label{sec:2.2}
\begin{figure}[t!] 
\center
\begin{tikzpicture}[scale=.8]
    \node[anchor=north,align=center] at (3.4,1) (IPOPT) { \footnotesize{$\mathrm{solve} \big \{ \mathrm{FOC}(\ldots) \big \}$} \\ \footnotesize{$\forall \; \bb{x}_t \subset {X}$}};	
    \begin{scope}[
            yshift=-83,every node/.append style={
            yslant=0.5,xslant=-1},yslant=0.5,xslant=-1
            ]fsca
        \fill[white,fill opacity=0.9] (0,0) rectangle (3,3);
        \draw[step=.5, black] (0,0) grid (3,3);
        \draw[black,very thick] (0,0) rectangle (3,3);
		\node[scale=0.6,fill=black] at (1,1) (X){};
        \draw[left] (IPOPT) edge[color=red,out=90,in=45,>->,line width=.5mm] (1,1) ;

    \end{scope} 
    \node at (1.8,-2.5) (P1) {\footnotesize{$\tilde{p}'$}};	
    \node at (-.4,-1.85)  (P1) {\footnotesize{$\bb{x}$}};	
    \node[red] at (0,1.5)  {\footnotesize{Approximation}};	
    \node[blue] at (7,1.5) {\footnotesize{Interpolation}};	   
    \begin{scope}[
    		shift={(7,0)},
            yshift=-83,every node/.append style={
            yslant=0.5,xslant=-1},yslant=0.5,xslant=-1
            ]
        \fill[white,fill opacity=0.9] (0,0) rectangle (3,3);
        \draw[step=.5, black] (0,0) grid (3,3); 
        \draw[black,very thick] (0,0) rectangle (3,3);
        
		\node[scale=0.6,fill=black] at (1.1,1.2)  (description) {};
		\node[scale=0.6,fill=black] at (.9,2.1)  (description) {};
		\node[scale=0.6,fill=black] at (1.2,2.9)  (description) {};
		\node[scale=0.6,fill=black] at (2.1,2.8)  (description) {};
		\node[scale=0.6,fill=black] at (2.0,1.3)  (description) {};
		\node[scale=0.6,fill=black] at (1.3,2.2)  (description) {};
		\node[scale=0.6,fill=black] at (2.1,2.5)  (description) {};
		\node[scale=0.6,fill=black] at (2.3,.7)  (description) {};	
		\draw (1.1,1.2)    edge[color=blue,out=45,in=30,->,line width=.5mm] (IPOPT);
		\draw (.9,2.1)    edge[color=blue,out=45,in=30,->,line width=.5mm] (IPOPT);
		\draw (1.2,2.9)  edge[color=blue,out=45,in=30,->,line width=.5mm] (IPOPT);
		\draw (2.1,2.8)  edge[color=blue,out=45,in=30,->,line width=.5mm] (IPOPT);
		\draw (2.0,1.3)  edge[color=blue,out=45,in=30,->,line width=.5mm] (IPOPT);
		\draw (1.3,2.2)    edge[color=blue,out=45,in=30,->,line width=.5mm] (IPOPT);
		\draw (2.1,2.5)  edge[color=blue,out=45,in=30,->,line width=.5mm] (IPOPT);		
		\draw (2.3,.7) edge[color=blue,out=45,in=30,->,line width=.5mm] (IPOPT);	\end{scope}
    \node at (9,-2.5)  (P2) {\footnotesize{$\tilde{p}$}};	
\end{tikzpicture}
\caption{
This figure is a visual representation of one step of the time iteration algorithm. 
We solve the first-order conditions (FOC) of the model for the state variable $\bb{x}_t$ in the updated policy function $\tilde{p}'$ (left), using the policy function from the previous time iteration step $\tilde{p}$ (right). 
}
\label{fig:1}
\end{figure}
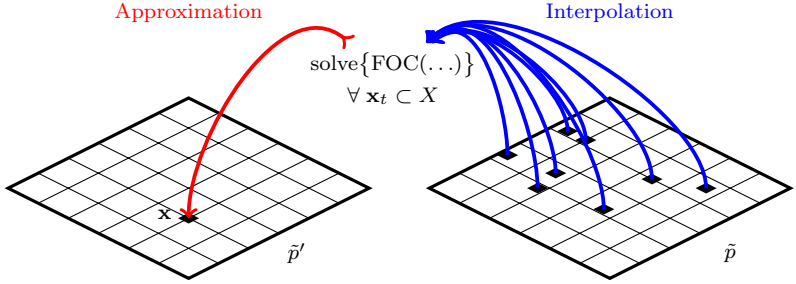

\begin{algorithm}[t!]
\caption{Time iteration algorithm.}
\label{alg:1}
\begin{algorithmic}[1]
{\fontsize{9}{11}\selectfont
\REQUIRE{$\tilde{p}'$, $\epsilon$.}
\REPEAT
	\STATE{$\tilde{p} \gets \tilde{p}'$}
	\FORALL{$\bb{x}_t \subset {X}$} 
		\STATE{$\tilde{p}'(\bb{x}_{t}) \gets  \underset{\bb{k}_{t+1},\bb{\mu}_{t},\lambda_t}{\mathrm{solve}} \big \{ \mathrm{FOC} \left (\bb{k}_{t+1},\bb{\mu}_{t},\lambda_{t}) | \bb{x}_{t},\tilde{p} \right)  \big \}, \;  $}
	\ENDFOR
\UNTIL{$\texttt{EulerEquationErrors}(\tilde{p}') <\epsilon$}	
\RETURN $\tilde{p}'$
}
\end{algorithmic}
\end{algorithm}
In this section, we briefly introduce the time iteration algorithm~\cite{coleman1990}. 
The latter solves for a recursive equilibrium of a dynamic economic model by guessing a policy function and iteratively updating it using the first-order equilibrium conditions of a given model such as those presented in Sections~\ref{sec:FOC_smooth} and~\ref{sec:FOC_nonsmooth}). For the sake of brevity, the discussion that follows only considers the non-smooth IRBC model. However, the algorithmic logic carries over to any model formulated analogously. Let $\bb{x}_t  \in {X} \subset \mathbb{R}^d$ denote again the state of the economy at discrete time $t$.\footnote{In practical applications, we restrict ourselves to the canonical domain, that is, ${\mathbf{x}}_t\in [0,1]^d$.}  
Recall that in the model under consideration, the economy is represented as a $(d=2N)$-dimensional state variable $\bb{x}_t=(\bb{a}_{t},\bb{k}_{t})$, and 
that the optimal policy function $p:X \to \mathbb{R}^{2N+1}$ maps the state variable to the capital choice, the KKT multipliers for the irreversible investments constraint, and the Lagrange multiplier for the aggregate resource constraint, that is, 
\begin{align}\label{eq:policy_it_2}
	 p(\bb{x}_t)=(\bb{k}_{t+1},\bb{\mu}_{t},\lambda_{t}),\; \text{and} \; p(\bb{x}_{t+1})=(\bb{k}_{t+2},\bb{\mu}_{t+1},\lambda_{t+1}).
\end{align}
The goal of the time iteration algorithm is to determine an approximate optimal policy function on the entire domain $X$, that is, to find a global solution to the problem stated in expression~\eqref{eq:social_planner} (see, e.g.,~\cite{Judd1998NumericalEconomicsb}, Section 17.8 for further details).
Such a policy function must satisfy the FOCs throughout the state space, that is,
\begin{align} \label{eq:FOC_solve}
	p(\bb{x}_{t}) =  \underset{\bb{k}_{t+1},\bb{\mu}_{t},\lambda_t}{\mathrm{solve}} \big \{ \mathrm{FOC} \left (\bb{k}_{t+1},\bb{\mu}_{t},\lambda_{t} | \bb{x}_{t},p \right)  \big \}, \;  \forall\, \bb{x}_t\; \in  {X}.
\end{align}
The time iteration procedure presented in Algorithm~\ref{alg:1} iteratively computes the approximate optimal policy function $\tilde{p}'$---as a numerical proxy to the true $p$---by using the previous policy iterate (or initial guess) $\tilde{p}$ to interpolate the values $\bb{k}_{t+2}$, $\bb{\mu}_{t+1}$, and $\lambda_{t+1}$ (cf. expression~\eqref{eq:policy_it_2}). The main components of the algorithm can be found in steps (3-5). In contrast to equation~\eqref{eq:FOC_solve}, we approximate $\tilde{p}'$ here; thus, the FOCs hold exactly (up to numerical precision) for the discrete grid points $\bb{x}_t \subset {X}$.
At each grid point where we solve~\eqref{eq:FOC_solve}, we require the solution for a system of $2N+1$ nonlinear equations, that is, for the FOCs presented in section~\ref{sec:FOC_nonsmooth}.
In order to solve this nonlinear set of equations, a large number of interpolated values from $\tilde{p}$ are required, as illustrated in Figure~\ref{fig:1}.
Finally, in step (6), the time iteration algorithm is stopped conditional on satisfying that the so-called \emph{Euler Equation Errors} are smaller than a given accuracy threshold. In economics, 
the latter criterion is a commonly used, unit-free measure of how accurately the equilibrium conditions are numerically satisfied (see, e.g.,~\cite{brummscheidegger_2017} Appendix C, or~\cite{FERNANDEZVILLAVERDE2016527} Section 7.2 for further details).
This process is repeated by swapping the policy function until the threshold $\epsilon$ is satisfied. 
The fundamental operations of Algorithm~\ref{alg:1}, which become computationally restrictive in a high-dimensional setting, are (i) generating the approximate policy function and (ii) costly interpolations from a policy function $\tilde{p}$ for solving the system of nonlinear equations.
Notice that the policy function we are interested in is high-dimensional and multivariate. In addition, due to the concavity assumptions on utility and production functions, the optimal policy $p$ will also be nonlinear (cf. Section~\ref{sec:FOC_nonsmooth}). Hence, approximating it only locally might provide misleading results. For such applications, we need a global solution and a performant interpolation method that approximates $p$ over the entire state space $X$. 
To do this efficiently, we will employ DDSG in combination with a hybrid parallelization scheme in Sections~\ref{sec:3} and~\ref{sec:4}, respectively. 

\section{Function approximation}  \label{sec:3}
The time iteration algorithm outlined in the previous section poses multiple computational challenges as it requires repeated function approximations and interpolation of a potentially high-dimensional policy function. 
This section outlines a function approximation method that addresses these concerns by using adaptive SGs and DD, which we discuss in Sections~\ref{sec:3.1} and \ref{sec:3.2}, respectively. 
With this background, we then combine the two numerical techniques in Section~\ref{sec:3.3} resulting in the DDSG approximation scheme for high-dimensional functions. 
We note that various forms of SG and DD exist; we will outline each method's key underlying concepts in a condensed fashion. 
For thorough introductions to SGs and DD, we refer to~\cite{Bungartz.Griebel:2004} and~\cite{genyuan,Rabitz1999GeneralRepresentations,Ma2010AnEquations}, respectively.

\subsection{Adaptive sparse grids}\label{sec:3.1} 
Let $f:\mathbf{x}\to\mathbb{R}^n$ where $\mathbf{x} \in [0,1]^d$, and $d$ in our case is the number of continuous state variables in the economic model of interest, a potentially large number. For the sake of brevity, we assume that the function value is zero on the domain's boundary.
This is not a necessary condition, and it can be easily changed by augmenting the basis function (see, e.g.,~\cite{IPDPS2018}). 
First, consider a one-dimensional domain, discretized with grid spacing $h_l=2^{-l}$.
The grid points are located at ${x}_{l,i}=i \cdot h_l$, where $i \in \{ 1,\ldots, 2^l \}$ and $l \in \mathbb{N}_{+}$ are the grid point indices and refinement level, respectively. 
Using the standard hat function, we define a family of univariate basis functions as 
\begin{align}  \label{eq:3.1} 
    \phi_{l,i}(x)=\max\left(1-\frac{1}{h_l}|x-{x}_{l,i}|,0 \right),
\end{align}
with support $[{x}_{l,i}-h_l,{x}_{l,i}+h_l]$.
The one-dimensional basis functions can be extended to a $d$-dimensional domain by introducing the multi-indices $\bb{i}=({i}_1,\ldots, {i}_d)$ and $\bb{l}=({l}_1,\ldots, {l}_d) \in \mathbb{N}_{+}^d $.  
The grid points are now denoted as $\bb{x_{l,i}}=({x}_{{l}_1,{i}_1}, \ldots ,{x}_{{l}_d,{i}_d})$,  and the corresponding $d$-linear hierarchical basis function is constructed by a tensor product, that is,
\begin{align} \label{eq:3.2} 
  \phi_{\mathbf{l},\mathbf{i}}(\bb{x}) = \prod_{j=1}^{d} \phi_{{l}_j,{i}_j}({x}_j).
\end{align}
Next, we introduce the hierarchical index-set $\bb{I}_{\bb{l}}$ and corresponding hierarchical subspace $W_{\bb{l}}$, which are given by: 
\begin{align}  \label{eq:3.3}
    \mathbf{I}_{\bb{l}}=\{\bb{i}: 0 < {i}_j < 2^{{l}_j}, {i}_j \; \text{odd}, 1 \leqslant j \leqslant d \},  
    \quad W_{\bb{l}}=\spn{ \phi_{\bb{l},\bb{i} }: \bb{i} \in \mathbf{I}_{\mathbf{l}}}.
\end{align}
Notice that the odd increments of ${i}_j$ result in the mutually disjoint support of the basis functions that cover the entire domain. 
For the space of piecewise linear functions,
\begin{align}\label{eq:3.4}
    V_\ell = \bigoplus_{ \norm{\mathbf{l}}{\infty} \leqslant \ell} W_{\mathbf{l}},
\end{align}
we can construct a corresponding equidistant Cartesian grid, also called a \emph{full grid}, with $M_{\ell}=2^{\ell}$ number of grid points in each dimension, where $\ell$ denotes the \emph{maximum refinement level}.
Approximations using the full grid will have an $L_2$ interpolation error of $\mathcal{O}(M_{\ell}^{-2})$ and number of grid points are of $\mathcal{O}(M_{\ell}^{d})$~\cite{Bungartz.Griebel:2004}.
It is clear that the full grid suffers from the curse of dimensionality and, thus, is not a scalable approach for high-dimensional function approximation. 

SGs can alleviate this issue; their underlying construction principle is to systematically eliminate those hierarchical increment spaces, which contribute only little to the overall quality of the approximation~\cite{Bungartz.Griebel:2004}. 
It can be shown that the SG space is given by 
\begin{align} \label{eq:3.5} 
    V^{\text{SG}}_\ell = \!\!\!\!\!\!\!\!\!\! \bigoplus_{\norm{\mathbf{l}}{1} \leqslant \ell+d-1} \!\!\!\!\!\!  W_{\mathbf{l}}.
\end{align} 
In contrast to the full grid space in~\eqref{eq:3.4}, where the maximum grid refinement levels in any dimension are restricted to $\ell$, now the sum is restricted.
The SG-based interpolation of a function $f$ at point $\bb{x}$, for a maximum refinement level $\ell$, can be uniquely expressed as 
\begin{align}\label{eq:3.6}
      \mathcal{I}_\ell f(\mathbf{x}) =  \sum_{ \norm{\mathbf{l}}{1} \leqslant \ell+d-1} \sum_{ \mathbf{i} \in \mathbf{I_l} }  \alpha_{\mathbf{l},\mathbf{i}}  \phi_{\mathbf{l},\mathbf{i}}(\mathbf{x}),
\end{align}
where the coefficients $\alpha_{\mathbf{l},\mathbf{i}} \in \mathbb{R}$, commonly referred to as the \emph{hierarchical surpluses}, can be readily computed (see, e.g.,~\cite{Bungartz2003MultivariateGrids} for details). 
Under certain technical assumptions---the function to be approximated needs to exhibit bounded second-order mixed derivatives---the SG approximation error is $\mathcal{O}(M^{-2}_{\ell}(\log M_{\ell})^{{d-1}})$, whereas the number of grid points grows as $\mathcal{O}(M_{\ell} \log(M_{\ell})^{d-1})$. Thus, the number of grid points in an SG is significantly reduced, whereas the error has only slightly deteriorated. Finally, quadrature on SGs can be performed very effectively, that is,  
\begin{align}\label{eq:3.7}
     \mathcal{Q}_\ell f =  \sum_{ \norm{\mathbf{l}}{1} \leqslant \ell+d-1} \sum_{ \mathbf{i} \in \mathbf{I_l} }  \alpha_{\mathbf{l},\mathbf{i}}   \int \phi_{\mathbf{l},\mathbf{i}}(\mathbf{x}) d\mathbf{x},
\end{align}
can be readily evaluated by integrating the basis functions in~\eqref{eq:3.2}---a key of SGs that we will leverage in the sections to come. In summary, the SG approximation method is a computationally efficient method for interpolation and or quadrature of sufficiently smooth functions on moderately high-dimensional domains. 

However, in situations where $f$ are highly nonlinear and show distinct local features, a high resolution level is required, but only in particular locations of the domain, which renders the ordinary SG inefficient. Adaptive SGs can cope with this issue to some extend. Unlike the ordinary SG, which has an a priori selection of grid points, adaptive SGs utilize an a posteriori refinement, which, based on a local error estimator, selects which grid points in the SG structure to be refined further~\cite{pflueger10spatially,Ma:2009:AHS:1514432.1514547}. For a predefined threshold $\epsilon_\gamma \in \mathbb{R}_+$ and
\begin{align}  \label{eq:3.8} 
  \gamma = \norm{\bb{\alpha}_{\bb{l},\bb{i}}}{\infty},
\end{align}
we deem a grid point to be significant if $\gamma > \epsilon_\gamma $. The refinement choice is governed by~\eqref{eq:3.8}; however, depending on the application, more sophisticated criteria may need to be imposed for an efficient refinement~\cite{scheidegger.treccani.20138}.
If a grid point is not accepted, all grid points that fall in its support will be excluded in the higher refinement level.

A parallel implementation of adaptive SGs is a computationally efficient technique to tackle nonlinear economic models of moderately high dimensions, say $d \leqslant 100$ in case the smooth IRBC models, or $d \leqslant 20$ in the non-smooth IRBC models~\cite{brummscheidegger_2017}.
However, when working with models of significantly higher complexity, such as those with state-space dimensions $>100$ or policy functions that exhibit high gradients, SGs, adaptive or not, are no longer a practical tool for function approximation.
In particular, the data structure becomes computationally demanding to operate on (see, e.g.,~\cite{ppopp027s-murarasu}).
Furthermore, increasing refinement levels to resolve non-smooth features in high-dimensional settings significantly increases the number of grid points, thus quickly rendering the approximation uncomputable~\cite{brummscheidegger_2017}.

\subsection{Dimensional decomposition}\label{sec:3.2}
In this section, we outline the DD approach that directly targets the curse of dimensionality by attempting to model the input-output behavior of a high-dimensional function as a sum of low-dimensional functions. As in the previous sections, we consider $f:\mathbf{x} \to \mathbb{R}$, where $\mathbf{x} \in [0,1]^d$.
Denote $\mathbf{u} \subseteq \mathcal{S}=\{1,2,\ldots,d\}$ as the \emph{component index}, and $f_{\mathbf{u}}:\mathbf{x_u} \to \mathbb{R}$ as the \emph{component function}, where $\bb{x}_{\bb{u}}$ is the vector comprising of the values $\mathbf{x}_i$ for $i \in \mathbf{u}$.  
The function $f(\bb{x})$ can be expressed as the hierarchical expansion 
\begin{align} \label{eq:ddsg2.2.1} 
  f(\mathbf{x}) &= f_{\emptyset}+\sum_{1\leqslant i \leqslant d} f_{i}({x}_i) +\sum_{1\leqslant i < j \leqslant d } f_{ij}({x}_i,{x}_j) + \ldots + f_{12\ldots d}({x}_1,{x}_2,\ldots, {x}_d ) 
\end{align}
where $f_{\emptyset}$ is a constant,  $f_i({x}_i)$ models the independent contribution, $f_{ij}({x}_i,{x}_j)$ the pairwise dependent contribution, and so on, up to the last term $f_{12\ldots d}({x}_1,{x}_2,\ldots, {x}_d)$, which accounts for the residual contributions.
In its complete form, the summation in~\eqref{eq:ddsg2.2.1} is exact, as the last term accounts for all contributions of all the input variables.
This representation is referred to as \emph{High-Dimensional Model Representation} (HDMR) and is a general method for a function decomposition that captures high-dimensional input-output system behavior~\cite{Rabitz1999GeneralRepresentations,genyuan,Hooker2007GeneralizedVariables,Li2012GeneralVariables}.
The summation in~\eqref{eq:ddsg2.2.1} can be compactly written as
\begin{align} \label{eq:ddsg2.2.2} 
  f(\bb{x}) &= \sum_{\bb{u} \subseteq \mathcal{S}} f_{\bb{u}}(\bb{x_u}).
\end{align}
The terms in the summation in \eqref{eq:ddsg2.2.2} are categorized by the \emph{expansion order} $k=\card{\mathbf{u}}$ or equivalently by the dimension of $\bb{x}_{\bb{u}}$.
Suppose this summation can be truncated to some maximum expansion order $\ekk \ll d$ without significant degradation in the approximation quality. In that case, one can reduce the overall dimensionality of the function. 
For example, if we consider a $100$-dimensional function, its first-order expansion $k=1$, will result in $100$ $1$-dimensional functions. 
With this said, selecting the appropriate $\ekk$ for the decomposition will undoubtedly depend on $f$. 

Two popular versions of HDMR are \emph{cut-} and \emph{ANOVA-HDMR}~\cite{RABITZ199911,genyuan}.\footnote{See \cite{Li2012GeneralVariables} for a high-level review of variations of cut-HDMR such RS-HDMR, mp-cut-HDMR, Mulitcut-HDMR, and lp-RS-hDMR.}
We will define the component function for both types of HDMR, but our focus will be on cut-HDMR.
As we will see, the cut-HDMR approach is a more fitting approach for our application due to its reliance on function evaluations as opposed to ANOVA-HDMR, which requires high-dimensional numerical integration.
\begin{figure}[t!]
    \centering
    \includegraphics[width=0.4\textwidth]{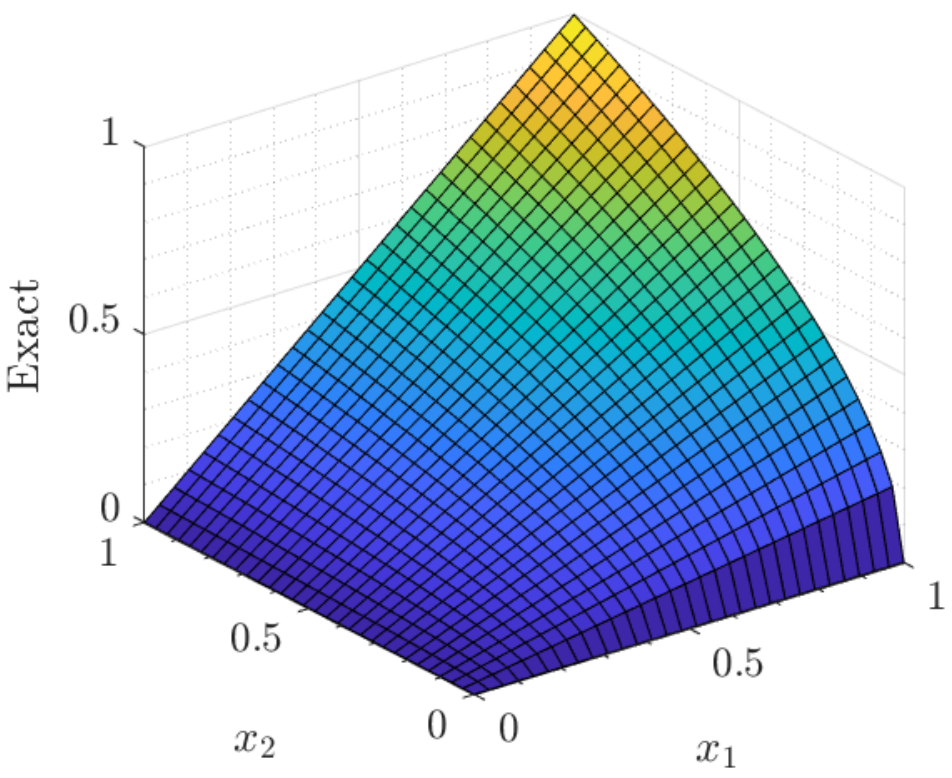} \quad\quad\quad\quad
    \includegraphics[width=0.4\textwidth]{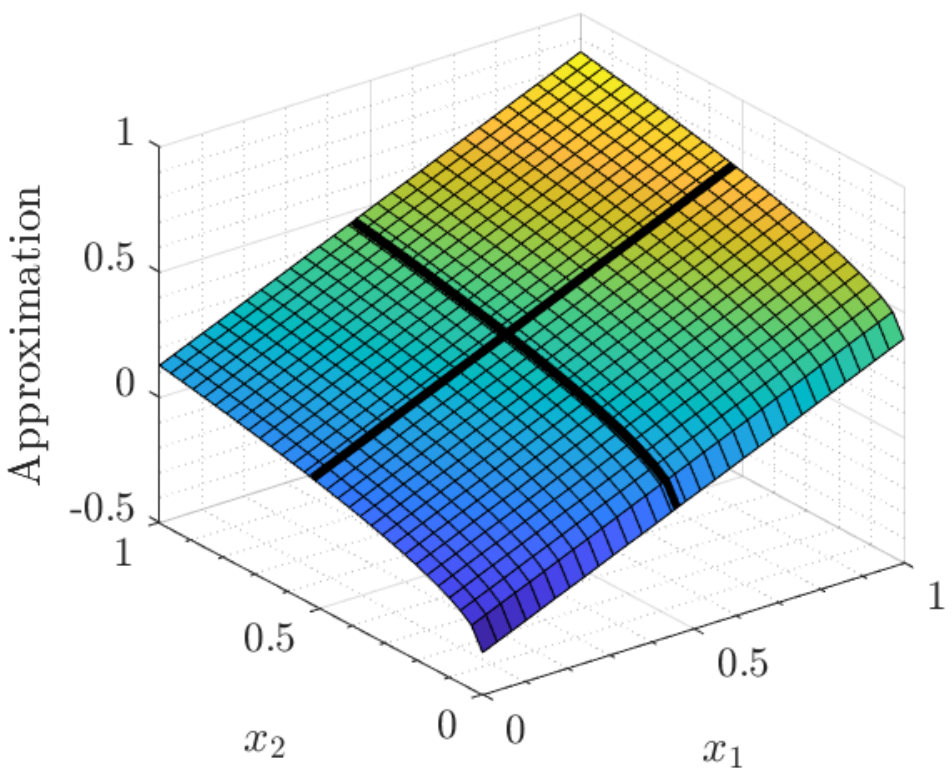}
     \caption{
     A plot of $f(\bb{x})=\bb{x}_1 \sqrt{\bb{x}_2}$ (left), and the corresponding first-order cut-HDMR approximation with its one-dimensional component functions shown with thick black lines (right).}
     \label{fig:cut_hdmr_examle}
\end{figure}
Let $w(\bb{x}) = \prod_{i=1}^d w_i({x}_i)$ be a product measure, with $w_{i}({x}_i)$ having a unit volume.
By sequentially ascending through the expansion orders, starting from the zeroth-order, the optimally and uniquely defined HDMR component function
\begin{align} \label{eq:ddsg2.2.4} 
   f_{\bb{u}}(\bb{x}_{\bb{u}}) =  &\argmin{g_{\bb{u}}} \int \left (  \sum_{\bb{u} \subseteq \mathcal{S}} g_{\mathbf{u}}(\mathbf{x_u}) - f(\mathbf{x}) \right )^2 w(\bb{x}) d \bb{x}, \\
     &\text{subject to} \;  \int g_\mathbf{u} (\mathbf{x_{u}}) w_{i}(\bb{x}_i) d {x}_i=0, \;\; \forall i \in \mathbf{u},\nonumber 
\end{align}
will only be dependent on lower-order component functions $f_{\bb{v}}(\bb{x}_{\bb{x}})$ for $\bb{v} \subset \bb{u}$.
This is particularly important, as the contrary would eliminate any reduction in dimensionality.
This attribute is a result of the imposed orthogonality condition in~\eqref{eq:ddsg2.2.4} ~\cite{Rabitz1999GeneralRepresentations,Hooker2007GeneralizedVariables}.
The cut-HDMR component functions are based on the Dirac measure, 
\begin{align}\label{eq:ddsg2.2.5}
 w(\bb{x}) d\bb{x} = \prod_{i=1}^d \delta({x}_i - \bar{x}_i)d {x}_i.  
\end{align}\noindent
where  $\bbb{x}=(\bar{x}_1, \ldots, \bar{x}_d)$ is a reference point in space, referred to as the \textit{anchor point}. 
Several techniques exist for the selection of the anchor point~\cite{GAO20103274}.
A straightforward approach is to simply choose the anchor point randomly in the high-dimensional space.
However, this can affect the approximation (i.e., for $\ekk<d$), and a careless selection may result in large errors (see, e.g.,~\cite{Wang2008OnRepresentation} for details on HDMR error analysis).
As noted in~\cite{Sobol2003TheoremsRepresentation} a suitable anchor point should satisfy
\begin{align}\label{eq:ddsg2.2.8}  
\displaystyle \min_{\bar{\mathbf{x}}} \left\|f(\bar{\mathbf{x}}) - \mathbb{E}[f(\mathbf{x})] \right\|_1.
\end{align}
An appropriate anchor point can be selected via sampling $\bbb{x}$ such that $f(\bbb{x})$ approximates the mean of the function over the domain (see, e.g.,~\cite{Ma2010AnEquations} for further details).
It is important to highlight that $\bbb{x}$ is not unique since a given function value could attain the mean value in several parts of the domain. 
The cut-HDMR component functions can be explicitly evaluated from~\eqref{eq:ddsg2.2.4} as a telescopic summation~\cite{Kuo2009OnFunctions}
\begin{align}\label{eq:ddsg2.2.6} 
    f^{\text{cut}}_{\mathbf{u}}(\mathbf{x_u}) = \sum_{\bb{v} \subseteq \bb{u}}(-1)^{\card{\bb{u}}-\card{\bb{v}}}f(\mathbf{x})\eval{\bb{x}}{\bar{\bb{x}} \backslash \bb{x_v}}, \quad \text{with}\, f_\emptyset &= f(\bar{\mathbf{x}}) .
\end{align}
We use the notation $\mathbf{x} = \bar{\mathbf{x}} \backslash \mathbf{x_v}$ to refer to assigning $\mathbf{x}$ the values of $\bar{\mathbf{x}}$ but excluding the indices of $\mathbf{v}$. 
For example, given $\mathbf{x}=({x}_1,{x}_2,{x}_3)$, then $\bbb{x} \backslash \bb{x}_{1,2}=({x}_1,{x}_2,\bar{x}_3)$. 
In Figure~\ref{fig:cut_hdmr_examle}, we depict an example $2$-dimensional function on the left panel and the first-order cut-HDMR approximation on the right panel. 

Using the Lebesgue measure, in place of the Dirac in~\eqref{eq:ddsg2.2.5} the ANOVA-HDMR decomposition is recovered~\cite{Holtz2010SparseInsurance}, with the component function 
\begin{align} \label{eq:ddsg2.2.7} 
    f^{\text{ANOVA}}_{\bb{u}}(\bb{x_u}) = \sum_{\bb{v} \subseteq \bb{u}}(-1)^{\card{\bb{u}}-\card{\bb{v}}} \int f(\bb{x}) d \bb{x}_{\mathcal{S} \backslash \bb{u} }, \quad \text{with}\, f_\emptyset &= \int f(\bb{x}) d \bb{x}.
\end{align}
Notice that on the $k$-th expansion order,~\eqref{eq:ddsg2.2.7} requires quadrature across $(d-k)$ dimensions.
Computationally, high-dimensional quadrature is a prohibitive operation subject to the same computational challenges we wish to address: the curse of dimensionality. 
In contrast, the cut-HDMR component functions are easily computed, requiring only function evaluation, which in turn for an arbitrary input $\bb{x}$, necessitates interpolation of $f(\bb{x})\eval{\bb{x}}{\bar{\bb{x}} \backslash \bb{x_v}}$.
As such, we proceed with adopting the cut-HDMR decomposition for which we will simply refer to as $f_{\bb{u}}(\bb{x}_{\bb{u}})$.
In principle, representing the cut-HDMR component functions is independent of the underlying numerical approach.
In practice, however, we will need a very efficient numerical approach since the component functions with high DD expansion orders will, to some extent, be exposed to the curse of dimensionality.

\subsection{Dimensional decomposition using adaptive sparse grids}\label{sec:3.3}
To approximate an economic policy function globally, we follow~\cite{Ma2010AnEquations} and~\cite{Yang2012AdaptiveFlows}, and embed adaptive SGs within DD to form the DDSG method.\footnote{The so-called \emph{SG combination technique} (see, e.g.,~\cite{Bungartz.Griebel:2004,Holtz2010SparseInsurance}, and references therein) provides an alternative construction of the basic DDSG method. However, we follow in our work the description of~\cite{Ma2010AnEquations}, where adaptive SGs are embedded within DD.} 
Utilizing adaptive SGs for the underlying numerical method has two desirable properties: (i) they can be applied to moderately high-dimensional component functions with non-smooth features, and (ii) quadrature operations can be carried out efficiently on them. 
Using the combined DDSG numerical approach, we can approximate the function $f$ as
\begin{align}\label{eq:3.16} %
  f(\mathbf{x}) &\approx \sum_{ \substack{ \mathbf{u} \subseteq \mathcal{S} \\  \card{\mathbf{u}} \leqslant \ekk  } } \sum_{\mathbf{v} \subseteq \mathbf{u}}(-1)^{\card{\mathbf{u}}-\card{\mathbf{v}}} \mathcal{I}_{\ell}f(\mathbf{x})\eval{\mathbf{x}}{\bar{\mathbf{x}} \backslash \mathbf{x_v}} ,   
\end{align}
which is a summation of $|\mathbf{v}|$-dimensional adaptive SGs, where here we truncate the DD maximum expansion order~$\ekk \ll d$. The accuracy of the DDSG approximation is dependent on the underlying function, SG approximation, maximum expansion order $\ekk$, and in turn, the number of component functions.\footnote{For details on the approximation error of the DDSG method, see~\cite{Ma2010AnEquations}.}

At a given expansion order, the DDSG component functions increase combinatorially.
The number of grid points in the approximation is the summation over the grid points of lower-dimensional SGs,
\begin{align}
    \sum_{k=1}^{\ekk} \card{V_{\ell,k}^{\text{SG}}} {{N}\choose{k}}  =
    \sum_{k=1}^{\ekk} \card{V_{\ell,k}^{\text{SG}}} \frac{d!}{(d-k)!k!}  
\end{align}
where $\card{V_{\ell,k}^{\text{SG}}}=\mathcal{O}( M_\ell \log(M_\ell)^{k-1})$ denotes the number of grid points for $k$-dimensional SG with maximum refinement level $\ell$ (see, e.g., Section~\ref{sec:3.1} for further details).\footnote{ 
The exact number of grid points for SG will depend on the type of SG used (see, e.g.,~
\cite{pflueger10spatially} for a comparison of different variations of SGs). 
}
For $\ekk \ll d$, the number grid points for DDSG is given by $\mathcal{O}( M_\ell \log(M_\ell)^{\ekk-1}  d^\ekk)$. 
In comparison to SG, we have removed the exponential dependence on $d$, but now instead, the problem is exponentially dependant on $\ekk$.
Even for moderately sized problems, the number of component functions corresponding to an adaptive SG can pose a computational challenge for values of $\ekk>1$.

To address this shortcoming, we outline two criteria that will efficiently truncate the expansion by ignoring its insignificant component functions.
Specifically, we will outline two DDSG adaptivity criteria for the relevance of (i) the proceeding expansion order and (ii) of each component function within an expansion order.
With these criteria, we can generate an adaptive variate of~\eqref{eq:3.16} by only proceeding to higher expansion orders when required and eliminating insignificant component functions within an expansion order.
We emphasize that the expansion criterion, or the active dimension selection for that matter, is not a measure of convergence of the DD expansion. 
These criteria provide an assessment of the importance of the current expansion or component function with respect to the previously computed values.
As such, the usage of the criteria, in particular with an aggressively high tolerance may lead to might lead to excessive truncation or pruning of the component functions.
As discussed in more detail in Section~\ref{sec:5.2}, these criteria can also be used as an analysis tool to understand the impact of the different component functions and or expansion orders on the overall approximation.

\subsubsection{Expansion criterion}
\begin{figure}[t] 
    \centering
    \includegraphics[width=0.4\textwidth]{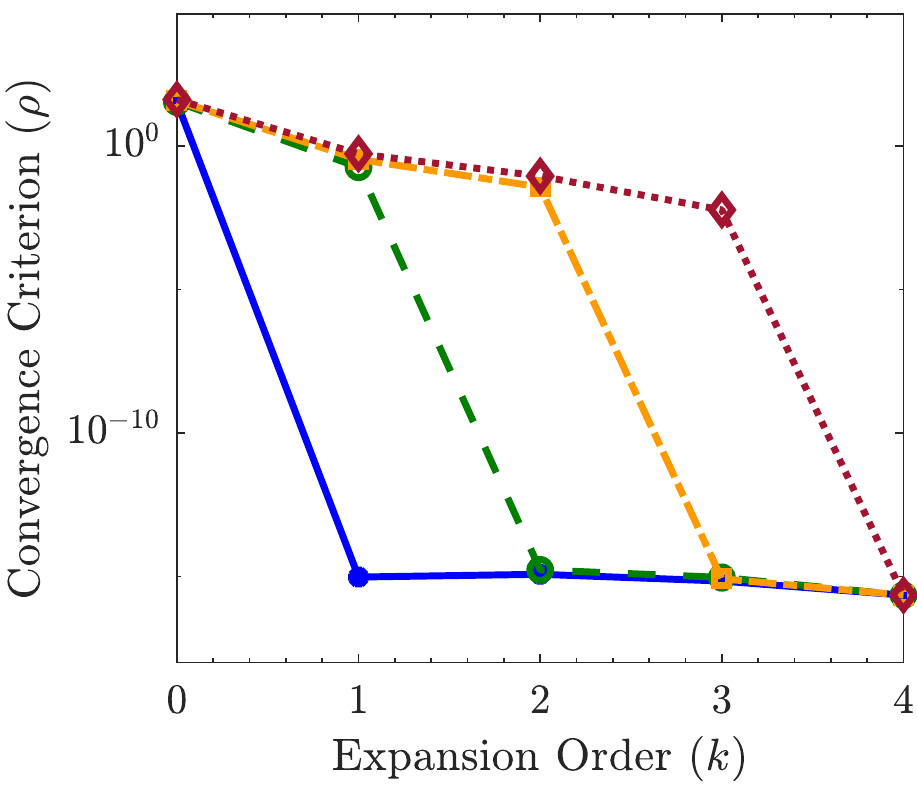} 
    \quad\quad\quad\quad
    \includegraphics[width=0.4\textwidth]{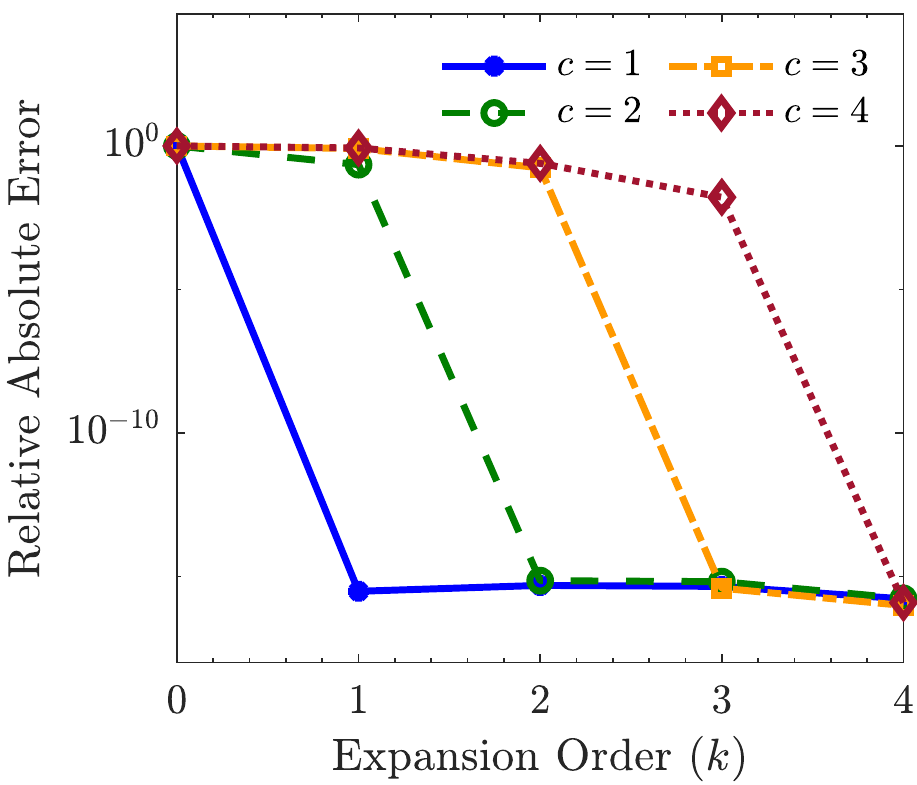}
    \caption{
    A plot of the DDSG expansion criterion (left) and the relative absolute error of the DDSG approximation (right) with respect to the expansion order $k$ are displayed.
    For both cases, the test function is a $4$-dimensional polynomial of degree $c$; i.e., ${f(\bb{x})=(\bb{x}_1+\ldots+\bb{x}_4)^c}$.
    The component functions are represented using SGs, the absolute relative error is evaluated using $10^3$ random samples, and $\bbb{x}=[0.5,0.5,0.5,0.5]$.
    }
    \label{fig:3}
    \vspace{-5px}
\end{figure}
The expansion criterion is the relative residual between two consecutive expansion orders $k$ and $k-1$. 
Given the current expansion order $k$, a predefined convergence threshold $\epsilon_\rho \in \mathbb{R}_+$, and a coefficient
\begin{align}\label{eq:3.18}
    \rho = \cfrac{ \norm{ \displaystyle \sum\limits_{ \card{\mathbf{u}} \leqslant k}  \mathcal{Q}_\ell f_\mathbf{u}(\mathbf{x_u}) d\mathbf{x} - \sum\limits_{ \card{\mathbf{u}} \leqslant k-1}    \mathcal{Q}_\ell f_\mathbf{u}(\mathbf{x_u}) d\mathbf{x} }{2}}{ \norm{ \displaystyle \sum\limits_{ \card{\mathbf{u}} \leqslant k-1}  \mathcal{Q}_\ell f_\mathbf{u}(\mathbf{x_u}) d\mathbf{x}   }{2}},
\end{align}\noindent
we justify the progression to the next expansion order $k+1$ if $ \rho>\epsilon_\rho $. 
The SG quadrature operation $\mathcal{Q}_\ell$ is given by~\eqref{eq:3.7}. 
With the assumption that $\ekk \ll d$, we can expect that the component function will not be high-dimensional, and thus, quadrature will not pose a computational bottleneck here. 

It is important to highlight that a truncation of~\eqref{eq:3.16} is not necessarily an approximation.
Indeed, the truncation can be error-free as long as the underlying function is additively separable up to the $\ekk$-th expansion order.
This feature is shown in Figure~\ref{fig:3}, where we display the value of $\rho$ and relative absolute error regarding the expansion order $k$ for a $4$-dimensional polynomial of varying degree $c$. Both $\rho$ and the relative absolute error reach machine precision when the expansion order is $k \geqslant c$.
We note that the expansion criterion is not a measure of convergence, as an increase in the expansion order does not necessarily translate into a decrease in the error.

\subsubsection{Active dimension selection criterion}
We look to identify insignificant component functions within an expansion order by using the active dimension selection criterion.
Here, we assume that the underlying function is not constant with respect to a single variable. 
Thus, we only focus on component function indices $\card{\mathbf{u}}\geqslant 2$. 
Given a predefined active component function threshold $\epsilon_\eta \in \mathbb{R}_+$ and coefficient 
\begin{align}\label{eq:3.19}
    \eta_\mathbf{u} = \cfrac{ \norm{ \displaystyle  \mathcal{Q}_\ell f_\mathbf{u}(\mathbf{x_u}) d\mathbf{x} }{2}}{ \norm{ \displaystyle \sum\limits_{ \mathbf{v}\subset \mathcal{S},\card{\mathbf{v}} \leqslant \card{\mathbf{u}}-1}  \mathcal{Q}_\ell f_\mathbf{v}(\mathbf{x_v}) d\mathbf{x}   }{2}},
\end{align}
we deem the component function index $\mathbf{u}$ as important if $ \eta_{\mathbf{u}}>\epsilon_\eta$.\footnote{Note that different criteria for~\eqref{eq:3.19} have been proposed in the literature so far (see, e.g.,~\cite{Yang2012AdaptiveFlows}).} 
Component indices that do not satisfy this condition and any superset of these indices are not computed. 
\begin{figure}[t!]
\centering
\begin{tikzpicture}
\begin{scope}[every node/.style={circle,thick,draw}]

    \node[rectangle] (123) at  (0,0)  {$f_{1,2,3}$};

    \node[rectangle] (12) at  (2,1)  {$f_{1,2}$};
    \node[rectangle] (23) at  (2,0)   {$f_{2,3}$};
    \node[rectangle] (13) at  (2,-1) {$f_{1,3}$};

    \node[rectangle] (1) at   (4,1) {$f_{1}$};
    \node[rectangle] (2) at   (4,0) {$f_{2}$} ;
    \node[rectangle] (3) at   (4,-1) {$f_{3}$} ;
    
    \node[rectangle] (0) at   (6,0) {$f_{\emptyset}$} ;
\end{scope}
\begin{scope}[every edge/.style={draw=black, thick}]
              
    \path[bend left ] [<-,>=latex] (123.north) edge (12.west);
    \path[          ] [<-,>=latex] (123.east) edge (23.west);
    \path[bend right] [<-,>=latex] (123.south) edge (13.west);

    \path[          ] [<-,>=latex] (12.east) edge (1.west);
    \path[          ] [<-,>=latex] (12.east) edge (2.west);    
    
    \path[          ] [<-,>=latex] (23.east) edge (2.west);
    \path[          ] [<-,>=latex] (23.east) edge (3.west);         
   
    \path[         ]  [<-,>=latex] (13.east) edge (1.west);
    \path[         ]  [<-,>=latex] (13.east) edge (3.west);     
 
    \path[bend left ] [<-,>=latex] (1.east) edge (0.north);
    \path[          ] [<-,>=latex] (2.east) edge (0.west);
    \path[bend right] [<-,>=latex] (3.east) edge (0.south);
\end{scope}
\end{tikzpicture}
\quad    
\begin{tikzpicture}
\begin{scope}[every node/.style={circle,thick,draw}]

    \node[rectangle] (23) at  (2,0)   {$f_{2,3}$};
    \node[rectangle] (13) at  (2,-1) {$f_{1,3}$};

    \node[rectangle] (1) at   (4,1) {$f_{1}$};
    \node[rectangle] (2) at   (4,0) {$f_{2}$} ;
    \node[rectangle] (3) at   (4,-1) {$f_{3}$} ;
    
    \node[rectangle] (0) at   (6,0) {$f_{\emptyset}$} ;
\end{scope}
\begin{scope}[every edge/.style={draw=black, thick}]

    \path[          ] [<-,>=latex] (23.east) edge (2.west);
    \path[          ] [<-,>=latex] (23.east) edge (3.west);         
   
    \path[          ] [<-,>=latex] (13.east) edge (1.west);
    \path[          ] [<-,>=latex] (13.east) edge (3.west);     
 
    \path[bend left ] [<-,>=latex] (1.east) edge (0.north);
    \path[          ] [<-,>=latex] (2.east) edge (0.west);
    \path[bend right] [<-,>=latex] (3.east) edge (0.south);
\end{scope}
\end{tikzpicture}
\caption{
A sketch of the component indices of a $3$-dimensional function (left), and the same with active dimension coefficient $\eta_{1,2}<\epsilon_\eta$ (right) are shown. 
All component function indices which form a superset of $\{1,2\}$ are ignored---that is, $\{1,2\}$ and $\{1,2,3\}$. 
In both cases, arrows signify the computational dependence, for example, $f_{1,3}$ is dependent on the values of $f_1$, $f_3$ and $f_\emptyset$.
}
\label{fig:4}
\vspace{-5px}
\end{figure}
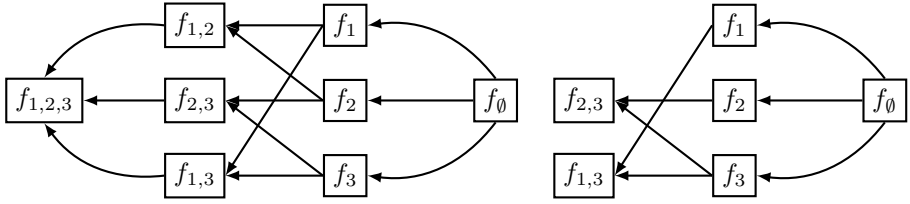
In Figure~\ref{fig:4}, we schematically depict the active dimension selection criterion on a three-dimensional function and the resulting ``pruning" effect on the combinatorial tree of component functions. 
In the left panel, we can see the component functions resulting from a full expansion. 
In the right panel, we display a scenario for which $\epsilon_{\eta}>\eta_{1,2}$ holds. 
In this case, the corresponding component index is removed from the computation, but also $\{1,2,3\}$ as $\{2,3\} \subset \{1,2,3\}$. 
Notice that the values from the quadrature operations in~\eqref{eq:3.19} can also be shared with~\eqref{eq:3.18}, and vice versa. 

\section{High-dimensional parallel time iteration framework}\label{sec:4}
This section introduces a performant framework for high-dimensional DDSG function interpolation and approximation, leveraging an optimized adaptive SG framework~\cite{IPDPS2018}. 
In Section~\ref{sec:4.1}, we describe a vectorized approach to DDSG interpolation, which allows for a performant, cache-efficient execution of function calls. 
Next, we outline in Section~\ref{sec:4.2} the parallelized DDSG time iteration framework for solving large-scale dynamic stochastic economic models.

\subsection{Vectorized DDSG interpolation} \label{sec:4.1}
In solving the system of nonlinear equations at a given point, the time iteration algorithm requires frequent interpolation on the policy functions $\tilde{p}$ from the previous iteration (see Section~\ref{sec:3.3} for further details).
These interpolations typically take up the majority---often far beyond 90\%~\cite{IPDPS2018}---of the computation time needed to solve the nonlinear set of equations. 
Therefore, the time-to-solution of the time iteration algorithm is highly sensitive to the performance of the DDSG interpolation function call. 
Direct implementation of~\eqref{eq:3.16} results in a massive number of repeated computations as many of the SG interpolants are identical. 
Consider for example, the component functions $f_{1,2}$ and $f_{1,2,3}$ both require the interpolation values of $\mathcal{I}_\ell f(\mathbf{x}) \eval{\mathbf{x}}{\bar{\mathbf{x}} \backslash \mathbf{x}_{1} }$ and $\mathcal{I}_\ell f(\mathbf{x}) \eval{\mathbf{x}}{\bar{\mathbf{x}} \backslash \mathbf{x}_{2} }$. 
Notice that the number of repeated computations increases nonlinearly with respect to the function's dimensionality and DDSG expansion order. 
Furthermore, a simple lookup-table approach would result in an erratic memory access pattern on a large array; thus, the computation would be plagued with cache misses.

We can eliminate all redundant SG interpolation and achieve an ideal access pattern without significant overhead in the memory footprint.
This is achieved by separating telescopic summation in~\eqref{eq:3.16} and storing the SG interpolation and coefficient values in two separate arrays  
\begin{equation}\label{eq:4.2}
\begin{rcases}
  \mathbf{a}_{\mathbf{i}} (\mathbf{x}) &=  \displaystyle \mathcal{I}_{\ell} f(\mathbf{x})\eval{\mathbf{x}}{\bar{\mathbf{x}} \backslash \mathbf{x_i}} ,\\
  \mathbf{b}_{\mathbf{i}} &= \displaystyle \sum_{\mathbf{u}\subseteq \mathcal{S}}  \sum_{\substack{ \mathbf{v} \subseteq \mathbf{u} \\ \mathbf{i}=\mathbf{v}} }  (-1)^{\card{\mathbf{u}}-\card{\mathbf{v}}}
\end{rcases}\quad  \; \{ \forall \, \mathbf{i}\subseteq \mathcal{S} : \card{\mathbf{i}} \leqslant \ekk \}.
\end{equation}
In addition, $\mathbf{b}$, as it is independent of $\mathbf{x}$, will only need to be computed once.\footnote{While the vectorized strategy is similar to the memoization technique~\cite{memoization}, we are also concerned with data contiguity for increased cache performance.}
Notice that for notation clarity, the formulation above does not consider the two DDSG adaptivity criteria described in Section~\ref{sec:3.3} and asserts the full expansion up to the maximum expansion order $\ekk$. 
The following section will provide an explicit algorithm describing how each component function is selected with the respective DDSG adaptivity criteria.
The DDSG interpolation function call now reduces to a desirable dot product $f(\mathbf{x}) \approx \mathbf{a}(\mathbf{x})^\top \mathbf{b}$ with a contiguous data access pattern. 

\subsection{Parallel DDSG time iteration algorithm}\label{sec:4.2}
We begin with the general description of the generic parallelized DDSG Algorithm~\ref{alg:2} and proceed to outline the inclusion of DDSG in the time iteration Algorithm~\ref{alg:1}.

The parallelized DDSG algorithm takes as input: the function $f$ to be approximated, maximum expansion order $\ekk$, expansion criterion tolerance $\epsilon_\rho$, active dimension selection tolerance $\epsilon_\eta$, anchor point $\bar{\mathbf{x}}$, maximum refinement level $\ell$, and adaptivity SG tolerance $\epsilon_\gamma$. 
We begin in step ($1$) with all compute instances initializing the empty, vectorized DDSG arrays defined in~\eqref{eq:4.2}, the zeroth-order component function $f_\emptyset=f(\bar{\mathbf{x}})$, and the \emph{reject index-set} $\bb{Z}=\emptyset$. 
The reject index-set collects all component function indices excluded from the DDSG expansion as per the active dimension selection criterion. 
We sequentially progress through the expansion orders in the body of the algorithm in steps ($2$--$20$). 
At expansion order $k$, the \emph{current order index-set} $\bb{C}$ is defined as the component indices of order $k$, which are not a superset of any of the indices in $\bb{Z}$. 
Expanding on the example in Figure~\ref{fig:4} with expansion order $k=3$, values of the reject and current order index set will be $\bb{Z}=\{\{1,2\}\}$ and $\bb{C}=\{\emptyset\}$ as $\{1,2,3\} \supset \{1,2\}$, respectively. 
Subsequently, at step($4$), we rebalance the compute resources evenly based on the current order index set, and the computation is carried out in parallel in steps ($5$--$14$). 
For each parallel SG interpolant, the quadrature value, and the active dimension selection coefficient $\eta_{\mathbf{i}}$ are computed for component index $\mathbf{i}$. 
Next, in steps ($9$--$13$), we employ the active dimension selection criterion for each component index. 
If the component function is accepted, we assign the DDSG vector arrays as defined in~\eqref{eq:4.2}. If not, the component index $\mathbf{i}$ is added to the rejected index set, and the SG interpolant and its quadrature values are discarded. 
Notice that each compute instance has a local or partial version of the computed variables within the parallel section of the algorithm. 
We perform the global synchronization upon exiting the parallel region at step($15$). 
In steps ($16$--$19$), with the globally available quadrature values available, we apply the DDSG expansion criterion~\eqref{eq:3.19}. 
If the threshold is reached, the routine terminates. Otherwise, we proceed to the next expansion order. 
The return values of the routine at step($21$) are the vectorized DDSG interpolation arrays. 
\begin{algorithm}[t]
{\fontsize{9}{11}\selectfont
\begin{algorithmic}[1]
\REQUIRE{$f,\ekk, \epsilon_\rho,\epsilon_\eta,\bar{\mathbf{x}},\ell,\epsilon_\gamma$}
\STATE{$ \texttt{initialize}:\; \{\mathbf{a},\mathbf{b} ,f_\mathbf{\emptyset}, \mathcal{Z} \} $}
\FOR{$k= 1$ \TO $\ekk$}
    \STATE{$\bb{C} \gets \{ \bb{C} \subseteq \mathcal{S}: \, \forall \; \mathbf{c} \in \bb{C}, \; \forall \; \mathbf{z} \in \bb{Z}, \;   \card{\mathbf{c}}=k, \;  \mathbf{c} \not\supset \mathbf{z}\}$}
    \STATE{$\texttt{load\_balance} \; \textit{given} \; \mathcal{C}$ }
    \FORALL{$\mathbf{i} \in \bb{C} \textbf{ parallel}$}
    	\STATE{$\texttt{compute}: \mathcal{I}_\ell f(\mathbf{x})\eval{\mathbf{x}}{\bar{\mathbf{x}} \backslash \mathbf{x_i}} 
			\quad\quad\quad\quad\;\, 
    		\triangleright \textit{Using SG adaptivity tolerance $\epsilon_\gamma$.}$}
    	\STATE{$\texttt{compute}: \mathcal{Q}_\ell f(\mathbf{x})\eval{\mathbf{x}}{\bar{\mathbf{x}} \backslash \mathbf{x_i}} 
			\quad\quad\quad\quad\,
    		\triangleright \textit{Using SG adaptivity tolerance $\epsilon_\gamma$.}$}
    	\STATE{$\texttt{compute}: \eta_{\mathbf{i}}$}
    	
    	\IF{$ \eta_{\mathbf{i}} \geq \epsilon_{\eta} $}
    	
    		\STATE{$\texttt{compute}: \{ \mathbf{a_i} (\mathbf{x}), \mathbf{b_i} \} \quad\quad\quad\quad\;\;\; \triangleright \textit{ As defined in~\eqref{eq:4.2}.}$}
    	\ELSE
    		\STATE{$ \bb{Z} \gets \{ \bb{Z}  \cup \mathbf{i} \}$}
    	\ENDIF
    \ENDFOR
    	
    \STATE{$\texttt{synchronize}: \{\bb{Z}, \mathcal{Q}_\ell f_{\ldots}, \mathbf{a}_{\ldots}, \mathbf{b}_{\ldots} \}$}
    \STATE{$\texttt{compute}: \{ \rho \}$}
    \IF{$\rho < \epsilon_\rho$}
        \STATE{$\texttt{break}$}
	\ENDIF
\ENDFOR   
\RETURN $\mathbf{a},\mathbf{b}$
\end{algorithmic}
}
\caption{Generic Parallel Adaptive DDSG Algorithm.}
\label{alg:2}
\end{algorithm}

\begin{figure}[t!]
\centering
    \includegraphics[width=.9\textwidth]{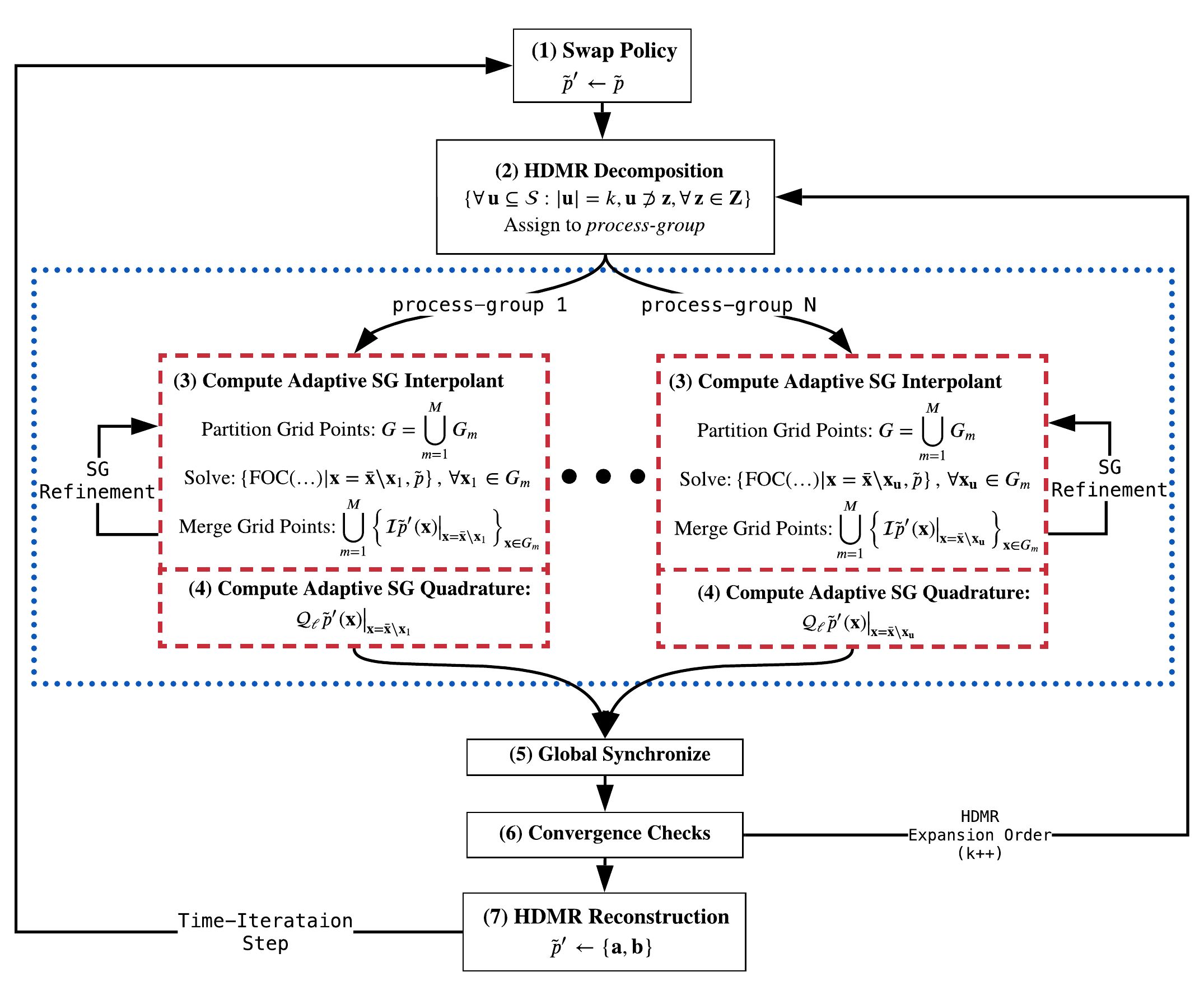}
    \caption{
The primary layer of parallelization (dashed blue lines) occurs in computing the HDMR component functions from steps ($2$--$5$). 
The secondary layer of parallelism (dotted red line) takes place while solving the system of the nonlinear equations at each SG grid point and carrying out quadrature operations, as shown in steps ($3$) and ($4$), respectively. 
    }
    \label{fig:update_prl_flow}
\end{figure}

In Figure~\ref{fig:update_prl_flow} we show the schematic for the parallel DDSG time iteration algorithm. 
In particular, we display two levels of parallelism: the first is based on distributed-memory (dashed blue lines), whereas the second relies on mainly share-memory\footnote{Both distributed- and shared-memory parallelization can be used in adaptive SG, though it is more effective to allocated distributed memory processes to the DD portion of the code \cite{Eftekhari:2017:PDD:3093172.3093234}.}(dotted red lines)\footnote{ \texttt{MPI} and Intel(R) \texttt{TBB} are used for distributed- and shared-memory parallelism, respectively.}. 
Starting in step($1$), we assign the newly computed policy function $\tilde{p}'$ as the current policy function $\tilde{p}$, or in the case of the initial step, we assign a random (guessed) policy function. 
Next in step($2$), we begin the HDMR decomposition starting from expansion order $k=1$ and evenly assign the component functions amongst groups of distributed-memory processes, referred to as a \emph{process-groups}. 
For each process group in step($3$), the respective SGs of the component functions are computed in parallel using a shared-memory approach. 
At a given refinement level, first, the SG grid points are evenly partitioned amongst the compute resources (threads), and second, at each grid point, we solve the first-order conditions\footnote{We use \texttt{IPOPT}~\cite{Wachter2005OnProgramming} for solving the FOCs.} noted in Section~\ref{sec:2.2}. 
We incrementally ascend through the SG refinement levels based on the adaptivity criterion described in Section~\ref{sec:3.1}. 
Next in step($4$), having computed the SG, we evaluate its quadrature for use in the DDSG adaptivity criteria noted in Section~\ref{sec:3.3}.
In step($5$),corresponding to step($15$) in Algorithm~\ref{alg:2}, we globally synchronize the distributed computations.
We proceed in step($6$) by checking the expansion criterion noted in \eqref{eq:3.18} and move to the next HDMR expansion order if required, that is, back to step ($2$).
Finally, in step ($7$), we reconstruct the DDSG policy function for the next iteration in step($1$).
With this parallelization approach, we expect that the parallel efficiency would be higher in the DD component, the primary layer, compared to the SG component, the secondary layer of the algorithm.
As highlighted in~\cite{Eftekhari:2017:PDD:3093172.3093234}, this is due to the unutilized computing resources in the SG algorithm at low refinement levels and also because of the repeated synchronization on increasing refinement levels. 
For example, in a one-dimensional SG with a maximum refinement level of $\ell=4$, we would have $1$,$2$,$4$, and $8$ grid points at each refinement level. 
Thus allocating $8$ cores for this computation would only achieve full utilization on the last refinement level. Furthermore, at each refinement level, we would require synchronization. 

\section{Results}\label{sec:5}
This section demonstrates the capabilities of the parallelized DDSG time iteration framework introduced in Section~\ref{sec:4}. 
We begin in Section~\ref{sec:5.1} with a set of basic performance tests for grid point reduction, vectorization performance, scalability, and speedup. 
In Section~\ref{sec:5.2}, we utilize DDSG as a tool to analyze both variates of the IRBC model described in Section~\ref{sec:2.1}. 
Using conclusions drawn from this analysis, in Section~\ref{sec:5.3}, we deploy the parallel DDSG time iteration framework to solve a set of large-scale smooth and non-smooth IRBC models.
We introduce the following naming standard for the parameterization of the DDSG routine: %
\begin{align}\nonumber
    \mbox{\(
    \text{DD}^{\epsilon_\eta}_{\ekk}\text{SG}_{\ell}^{\epsilon_\gamma}
    \)}. 
\end{align}
It should be assumed that $\epsilon_\rho=\epsilon_\eta$ (see Section~\ref{sec:3.3} for definitions of $\epsilon_\rho$ and $\epsilon_\eta$) unless otherwise noted. 
If a nonadaptive variate of the DDSG method is used, the values of $\epsilon_\eta$ and $\epsilon_\gamma$ are omitted. 
To measure our scheme's convergence, we follow the previous literature and report the \emph{Euler Equation Errors} (see~\cite{brummscheidegger_2017}, Appendix C for details). 
A total of $10,000$ error samples are gathered for which the maximum (Maximum Euler Error) and average (Average Euler Error) are reported in $\log_{10}$ scale. 
In all test cases, the reported targeted Average and Maximum Euler errors in our approximate solutions align with the current literature. All experiments were conducted on the Piz Daint supercomputer at CSCS (Cray XC$50$ with $12$-cores and $64$GB of memory per node). 

\subsection{Unit Tests}\label{sec:5.1}
We now outline a set of unit test results to highlight the reduction in the number of grid points, the performance of the vectorized interpolation, the parallel scalability, and the speedup of the framework with respect to the state-of-the-art SG framework.  

\subsubsection{Grid point reduction}\label{sec:5.1.1}
\begin{figure}[t]
    \centering
        \includegraphics[width=0.4\textwidth]{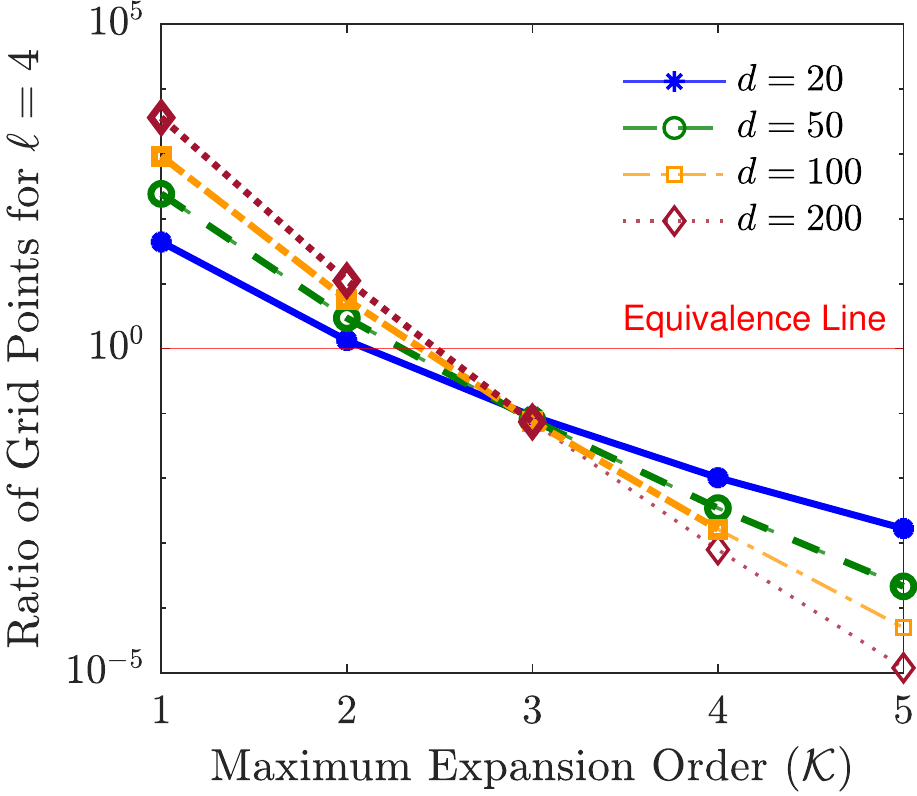} \quad\quad\quad\quad
    \includegraphics[width=0.4\textwidth]{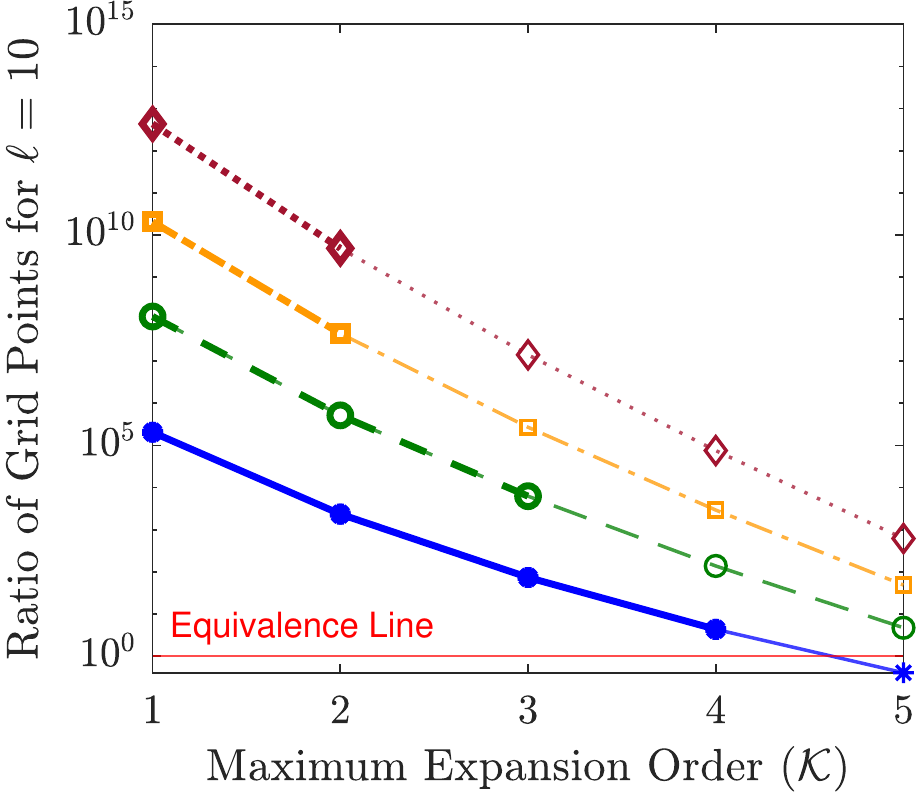}
       \caption{
       A plot of the ratio of SG to DDSG grid points for a maximum refinement level $\ell=4$ (left), and $10$ (right) with respect to the maximum expansion order $\ekk$ at varying dimension.
       In both figures, the thicker lines represent data points for which the number of DDSG grid points is $<10^9$.
       }
    \label{fig:6}
\end{figure}
In Figure~\ref{fig:6}, we show the SG to DDSG grid point ratios with respect to the maximum expansion order $\ekk$ for various function dimensionalities.
We refer the reader to Section~6 for further details on the adopted notation standard $\text{DD}^{\epsilon_\eta}_{\ekk}\text{SG}_{\ell}^{\epsilon_\gamma}$.
The red line in each graph denotes a ratio of one---that is to say, where the number of grid points for DDSG and SG are equivalent.
The plot lines not marked with thick lines are where using DDSG would result in a number of grid points $>10^9$. 
In these conditions, memory issues would render DDSG, or even SG, inoperable.
The two plots presented correspond to two scenarios, one where the underlying function exhibits relatively smooth dynamics (lower maximum refinement levels) and the other where one wishes to resolve non-smooth futures (higher maximum refinement levels). 
For a lower refinement level $\ell=4$, shown in the left panel, DDSG provides a reduction in grid points up to a maximum expansion order $\ekk=2$. 
At a maximum expansion order of $\ekk=1$, we can see orders of magnitude in grid point reduction.
This trend is further exaggerated when we require higher SG refinement levels. 
We show the same test in the right panel but with $\ell=10$. 
Here there is a reduction in grid points up to an expansion order of $\ekk=5$. 
However, at such high expansion orders and refinement levels, the overall number of grid points is much too high for practical usage. 

\subsubsection{Function call performance}\label{sec:5.1.2}
 \begin{figure}[t] 
    \centering
    \centering
    \includegraphics[width=0.4\textwidth]{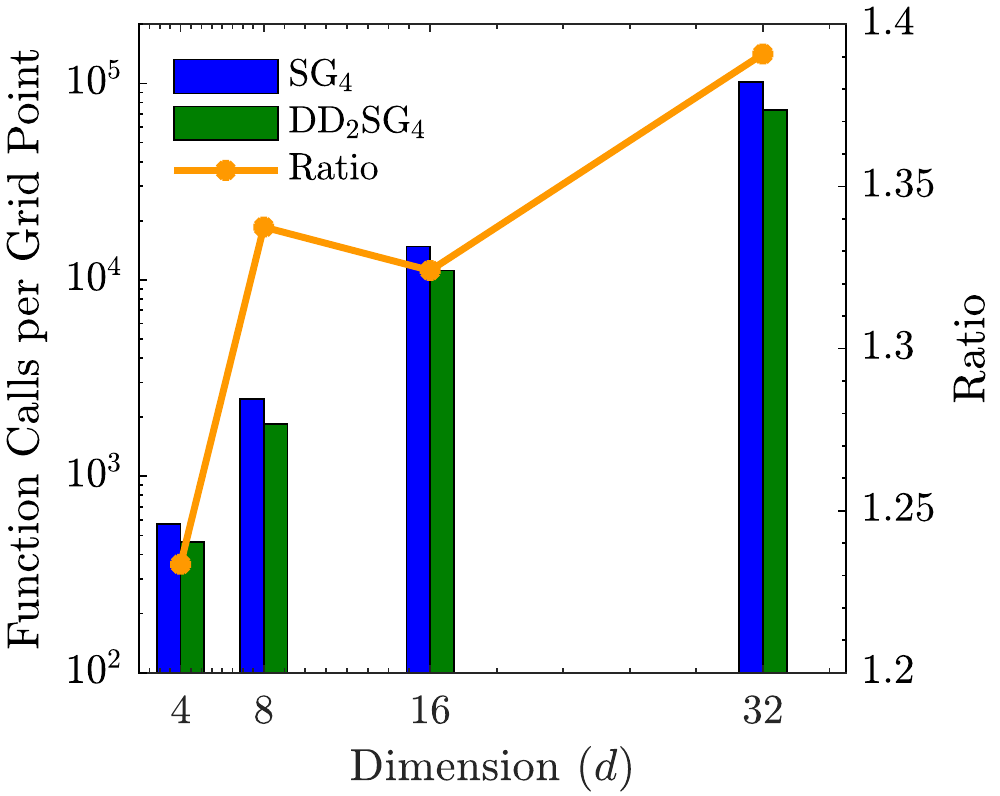} \quad\quad\quad\quad
    \includegraphics[width=0.4\textwidth]{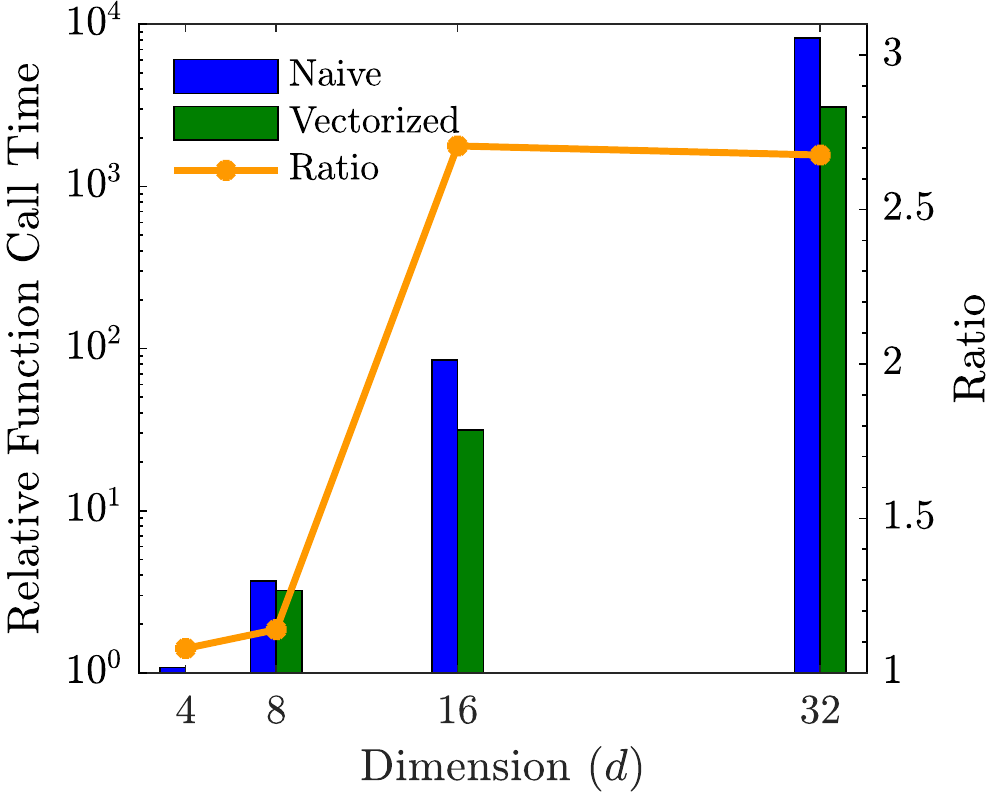}
    \caption{A plot of the number of function calls per grid point for SG$_4$ and DD$_2$SG$_4$ (left), and relative function call time of the DDSG naive and vectorized DDSG interpolation (right), with respect to varying dimensions for the smooth IRBC model.}
    \label{fig:8}
\end{figure}
The solution time for the first-order conditions of the IRBC models outlined in Sections~\ref{sec:FOC_smooth} and~\ref{sec:FOC_nonsmooth} is heavily dependent on both the number of optimizer function invocations per grid point and the time required for the function invocation. 
In Figure~\ref{fig:8}, we present the results for one step of the time iteration using SG$_4$ and DD$_2$SG$_4$. 
In all tests, the optimizer is set with a termination tolerance of $10^{-4}$.
In the left panel, we can see that DDSG provides roughly $30\%$ reduction in the number of function calls compared to SG. 
In the right panel, we see the relative compute times of the naive and vectorized approach (see Section~\ref{sec:4.1}) for DDSG interpolation. 
Here the naive approach is the evaluation of \eqref{eq:3.16} directly.
We can see that the vectorized approach provides a speedup of roughly $2.8$ times that of the naive implementation. 

\subsubsection{Scalability}\label{sec:5.1.3}
\begin{figure}[t]
\centering
        \includegraphics[width=0.4\textwidth]{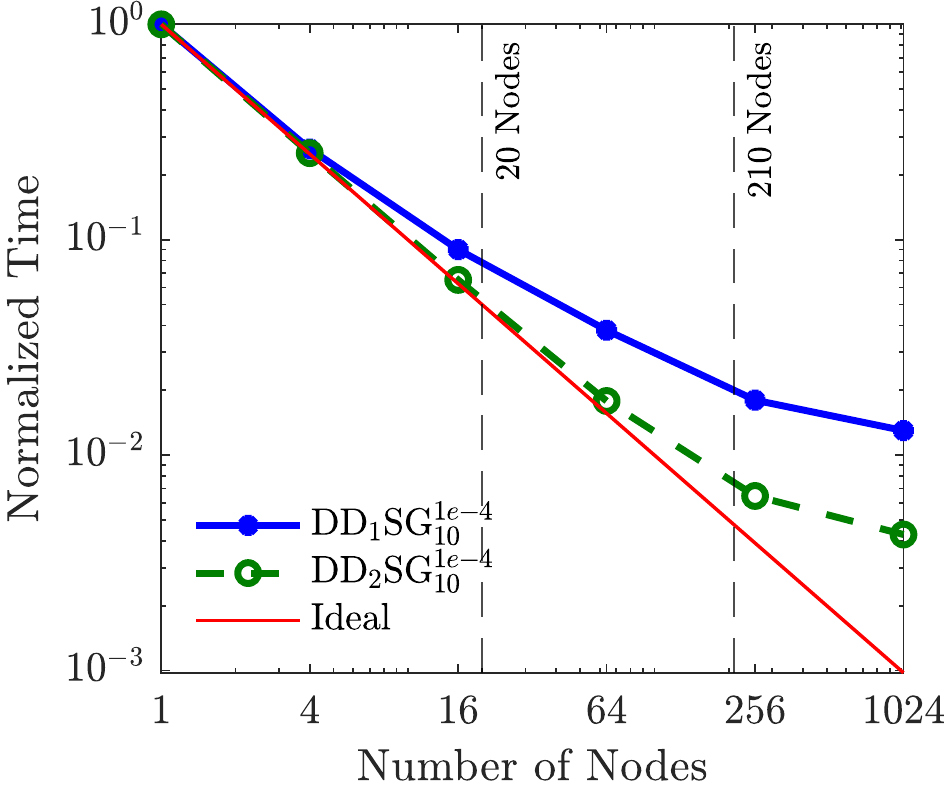} \quad\quad\quad\quad
    \includegraphics[width=0.4\textwidth]{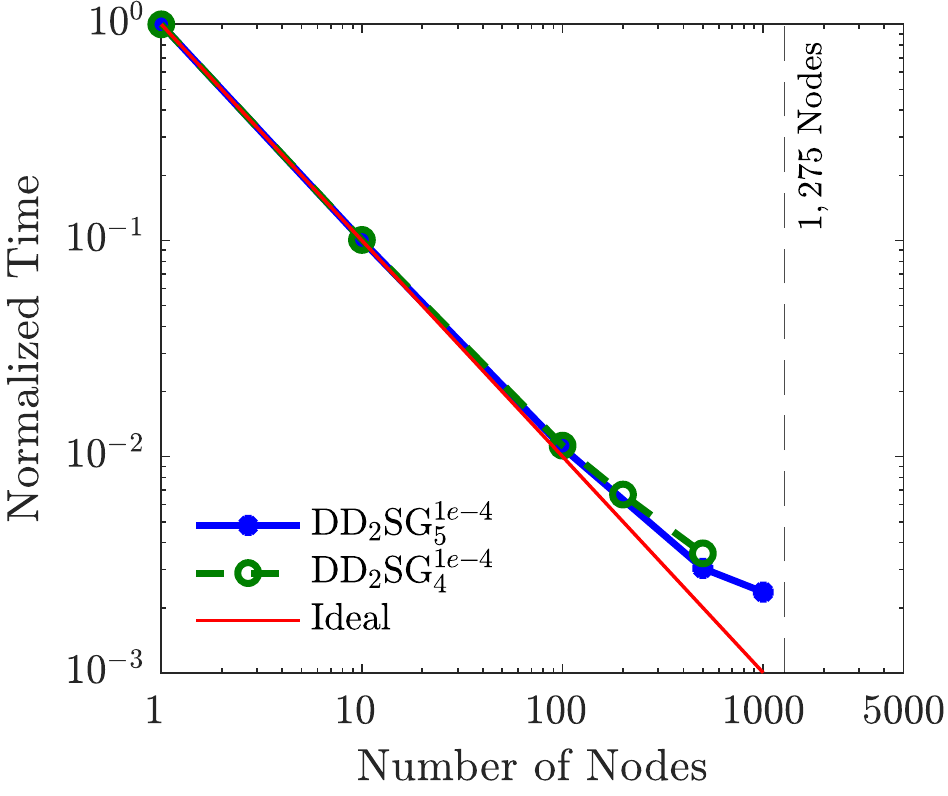}
    \caption{
    Normalized compute time of a $20$- (left), and a $50$-dimensional model (right) taking one time iteration step using DDSG with expansion orders $1$ and $2$, and maximum refinement levels $4$ and $5$, respectively.  
    }
\label{fig:scaling}
\end{figure}
In Figure~\ref{fig:scaling}, we show the normalized compute time for one step of the time iteration for a $20$ and $50$-dimensional IRBC model.
As noted in \ref{sec:4.2} and described in detail in \cite{Eftekhari:2017:PDD:3093172.3093234}, we expect the primary layer of parallelization, the DD component of the DDSG algorithm, to have better parallel efficacy than the SG component, that is, the secondary layer. 
The left panel shows numerical results for a $20$-dimensional model with a fixed maximum refinement level of $10$ and a varying maximum expansion order $1$ and $2$. 
Here, we have $20$ and $210$ component functions for the respective maximum expansion orders.
We can see almost ideal strong scaling up to a number of nodes equal to the number of component functions for both tests.
After this point, additional parallelism is taken from the secondary, less efficient layer of parallelism in the SG algorithm.
In the right panel, we use a $50$-dimensional model with a fixed maximum expansion order of $2$ and varying maximum refinement levels of $4$ and $5$. 
In contrast to previous test cases, we have $1,275$ component functions for both tests, which is larger than the maximum number of nodes. 
Here we can observe almost ideal strong scaling up to $1,000$ nodes, which is the same for both tests.
Furthermore, we see that the difference in the SG maximum refinement level slightly affects the parallel performance, with higher maximum refinement levels providing a marginal advantage in scalability.

\subsubsection{Speedup}\label{sec:5.1.4}
\begin{figure}[t]
\centering
        \includegraphics[width=0.4\textwidth]{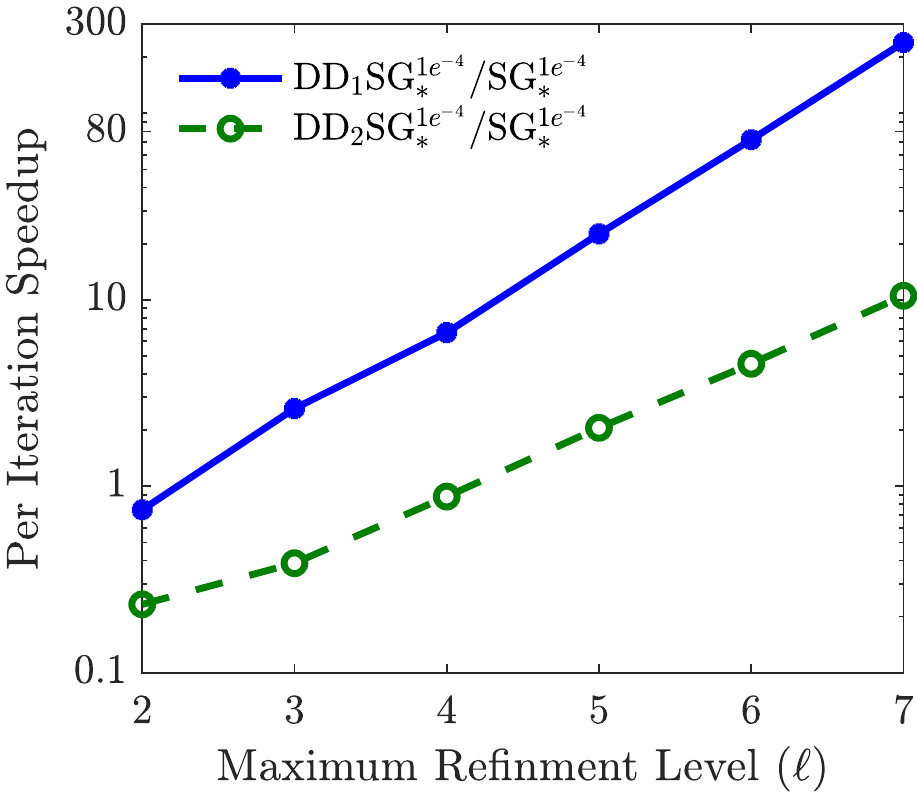} \quad\quad\quad\quad
    \includegraphics[width=0.4\textwidth]{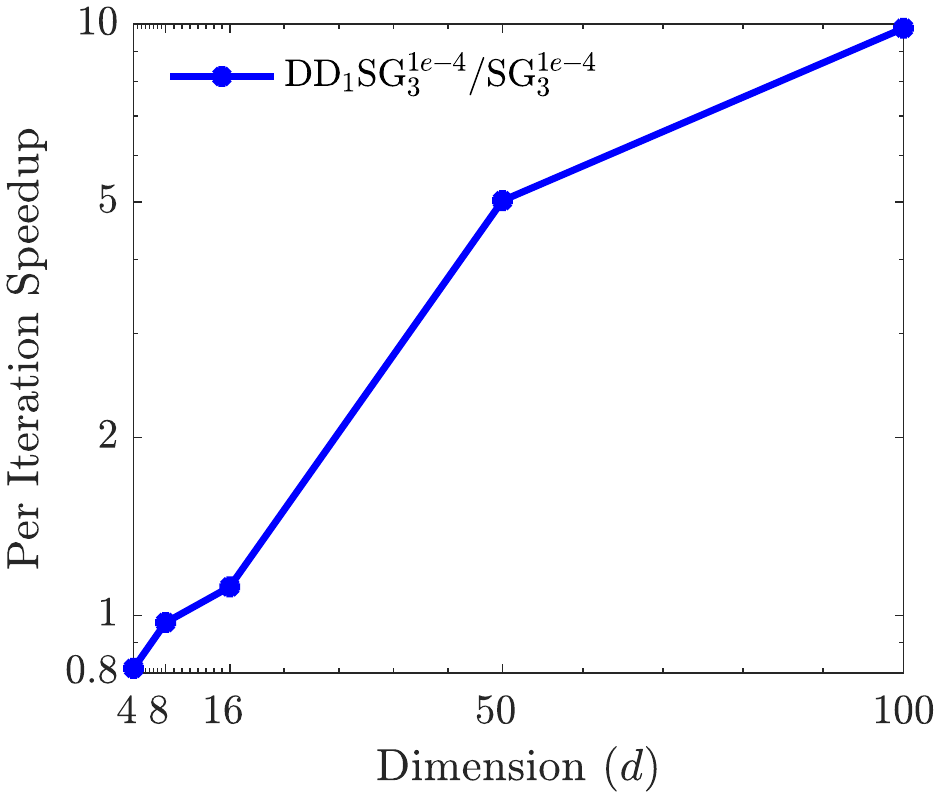}
   \caption{
   The plot of the speedup of DDSG over SG for a $8$-dimensional smooth and non-smooth IRBC model with respect to maximum refinement levels (left), and varying dimensionality (right).
   The non-smooth IRBC model represents the green dashed line for which we used $\ekk=2$.
   }
   \vspace{-5px}
\label{fig:speedup}
\end{figure}
%
In Figure~\ref{fig:speedup}, we display the speedup of the parallelized DDSG time iteration framework with respect to an SG version. 
We show a single time iteration step of an $8$-dimensional model for DDSG at maximum expansion order $1$ and $2$ at varying refinement levels in the left panel. 
This test is done on a single node using shared-memory parallelism, and both DDSG and SG approximation methods use the same adaptive coefficient. 
At refinement level $7$, for the respective tests, we can see that the DDSG approach provides $240$ and $10$ times faster runtimes than the SG approach. 
In the right panel, we show the time-to-solution for a single DDSG time iteration step at a maximum expansion order of $1$, maximum refinement level $3$, and varying dimensions for the smooth IRBC model. 
The tests are deployed on a number of nodes equal to that of the dimensionality of the model. 
The DDSG routine provides a speedup of up to $10$ times over SG implementation. 
The reason for this speedup is primarily due to DDSG operating on $100$ one-dimensional SGs while the SG time iteration framework operates on a $100$-dimensional SG. 

\subsection{Model analysis} \label{sec:5.2}
\begin{figure}[t]
    \centering
        \includegraphics[width=0.4\textwidth]{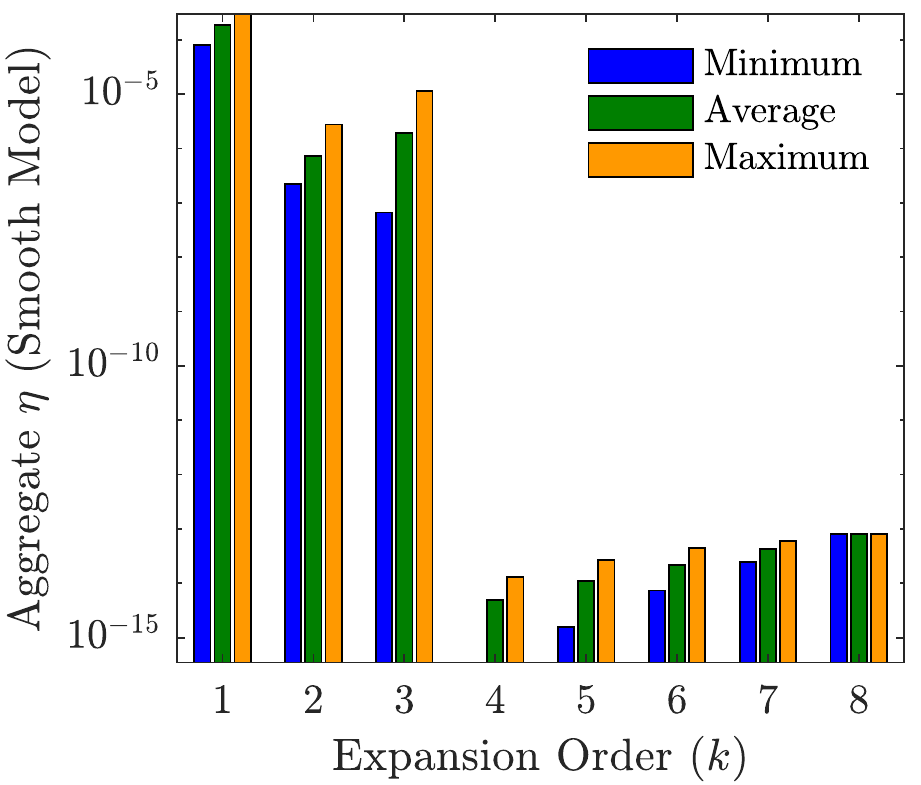} \quad\quad\quad\quad
    \includegraphics[width=0.4\textwidth]{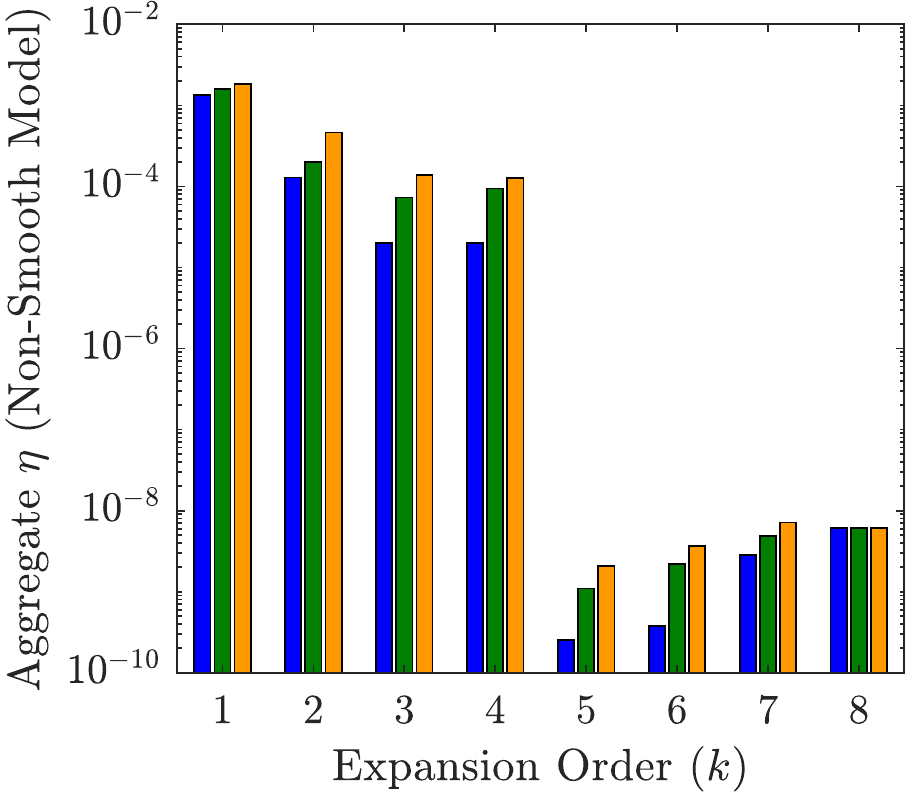}
    \caption{
    Aggregate minimum, average, and maximum values of the active dimensional selection criterion for the smooth (left), and the non-smooth (right) $8$-dimensional IRBC models.
    }
    \vspace{-10px}
\label{fig:9}
    \end{figure}

We now use the active dimension selection criterion outlined in Section~\ref{sec:3.3} as an analysis tool to assess the significance of the component functions of the policy function. 
In Figure~\ref{fig:9}, we show the aggregate minimum, average, and maximum values of $\eta_{\mathbf{u}}$ for the $8$-dimensional smooth and non-smooth models, at varying expansion orders. 
For a given expansion order, an increase in variability between the maximum and minimum values of $\eta_{\mathbf{u}}$ signifies that active dimension selection could be effective in selecting only a subset of the component functions. 
In contrast, if both the maximum and minimum values are equivalent to the average, then we would conclude that no component function could be considered to be more significant than the other. 
In such scenarios, active dimension selection would not be an effective way to reduce the computational cost of approximating the respective policy functions. 
%
In the left panel of Figure~\ref{fig:9}, we can see that the smooth model's policy function only significantly impacts up to the third expansion order in the left panel.
The second and third-order terms are approximately $100$ times less significant compared to the first-order component functions. 
Our experiments show that component functions with $\eta_{\mathbf{u}}<10^{\text{-}4}$ do not play a significant role in the approximation of the policy function. 
Thus, they can be eliminated without significantly degrading the overall approximation quality. 
Notice that using $\epsilon_{\eta}=\epsilon_{\rho}=10^{\text{-}4}$, the active dimensional selection criterion, and the expansion criterion would truncate the expansion at the first DDSG expansion order. 
In the right panel of Figure~\ref{fig:9}, we see the same analysis for the non-smooth IRBC model, whereas in this case, the component functions show significance up to the fourth-order expansion terms. 
Here we can see that both the first and second-order component functions are required while the majority of the third and fourth-order component functions fall below the $10^{\text{-}4}$ threshold. 
With this analysis, we proceed with our experiments using $\ekk=1$ and $2$ for the smooth and non-smooth IRBC models, respectively, and $\epsilon_\eta=\epsilon_\rho=10^{\text{-}4}$ for both cases.

\begin{figure}[t]
\centering
\;
     \includegraphics[width=0.4\textwidth]{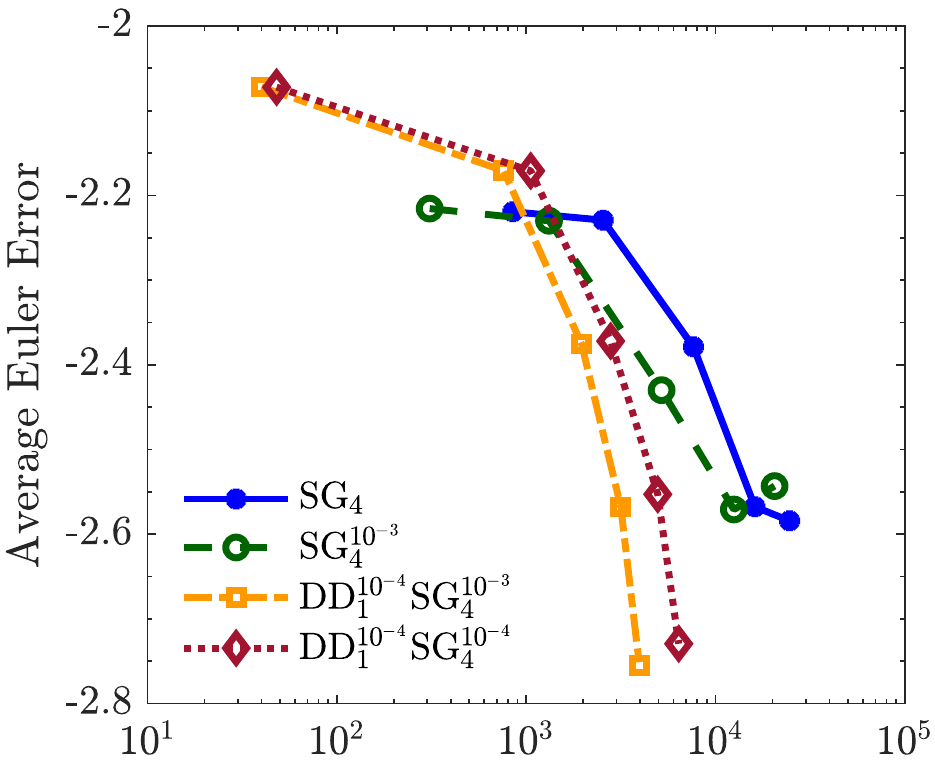}  \quad\quad\quad\quad
     \includegraphics[width=0.4\textwidth]{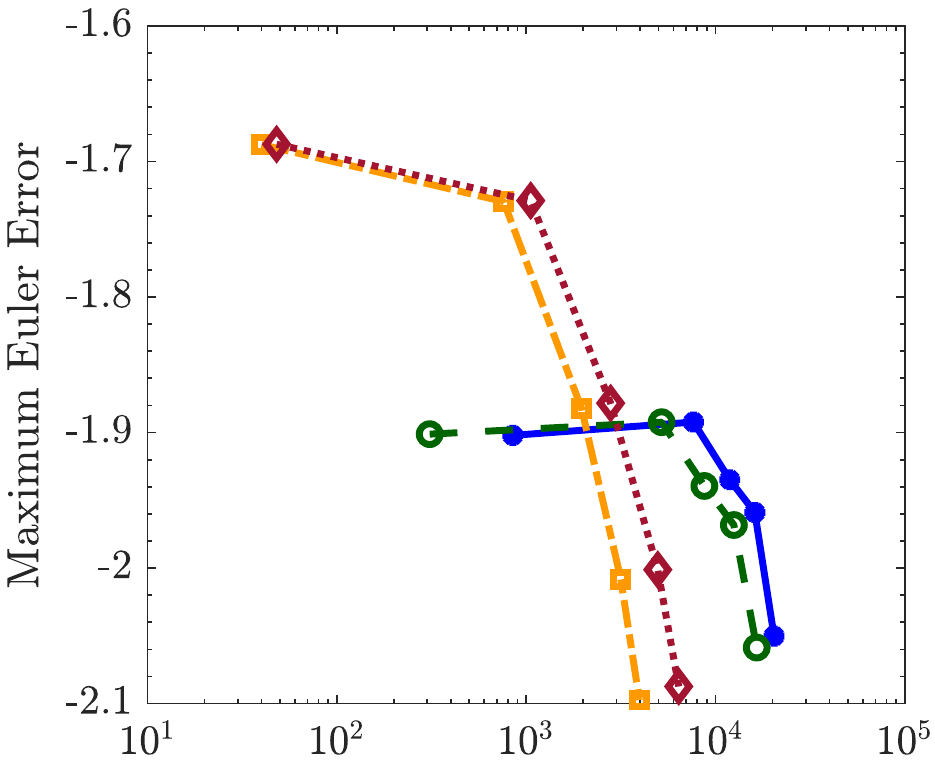}
     \newline
     \includegraphics[width=0.4\textwidth]{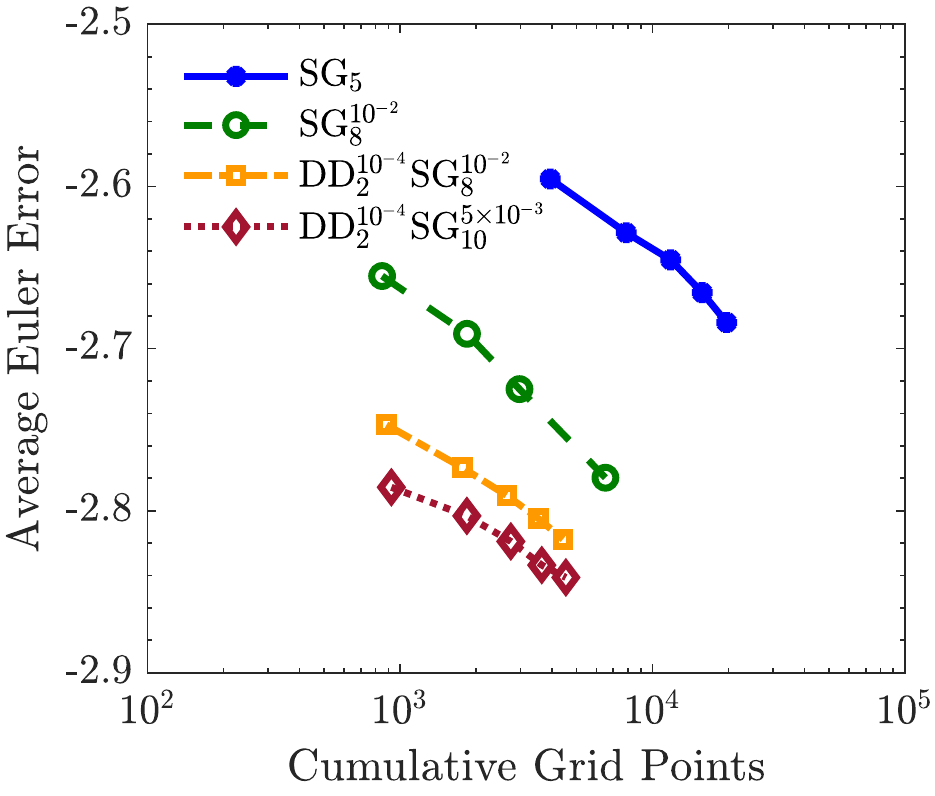}
     \quad\quad\quad\quad
     \includegraphics[width=0.4\textwidth]{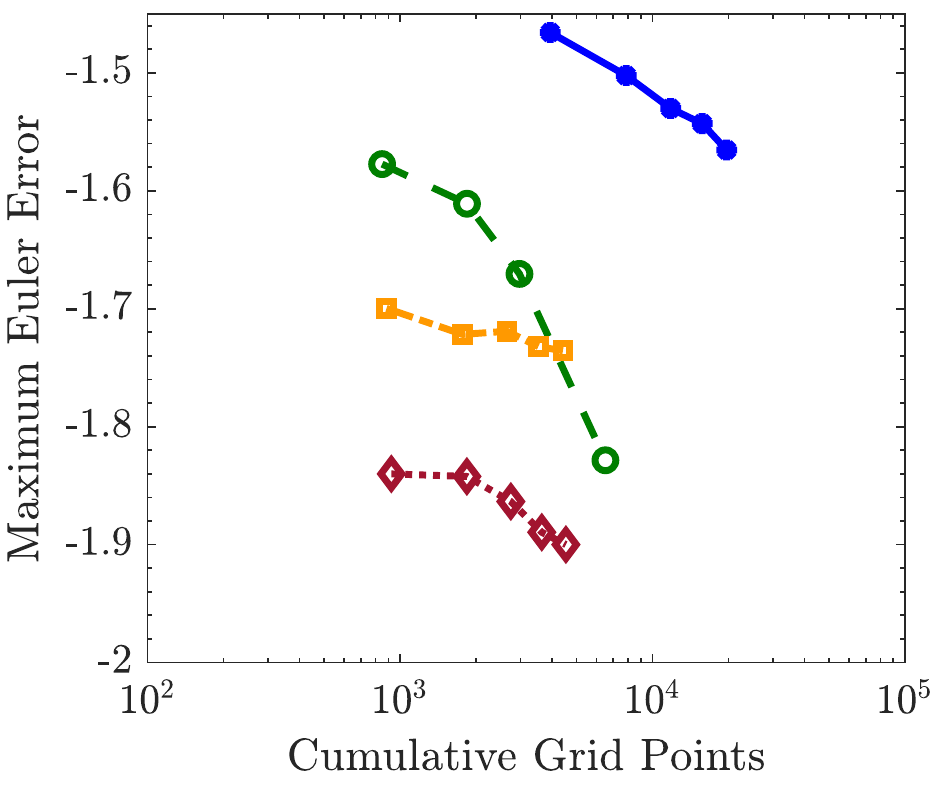}
    \caption{
    Average and maximum Euler error with respect to the cumulative number of grid points for an $8$-dimensional smooth (top panel) and non-smooth (bottom panel) IRBC model. 
    Notice that each data point represents a time iteration step.
    }
 \label{fig:10}
    \end{figure}
%
We now look at the effect of these parameters on the convergence trajectories. 
In the top panel of Figure~\ref{fig:10}, we show the average and maximum Euler errors for the smooth IRBC model using SG, adaptive SG, and the DDSG time iteration algorithm. 
The horizontal axis corresponds to the cumulative number of grid points evaluated for several time iteration steps. 
For DDSG to be a superior approximation method than SG and adaptive SG, we should attain smaller Euler errors for the same number of grid points. 
We test our DDSG implementation with $\ekk=1$ at $\ell=4$, using $\epsilon_\gamma=10^{\text{-}3}$ and $10^{\text{-}4}$. 
The observed average and maximum Euler errors begin relatively high in both configurations but quickly decrease beyond classical and adaptive SG. 
Notice that DDSG with $\epsilon_\gamma=10^{\text{-}4}$ does not improve the convergence rate in comparison to DDSG with $\epsilon_\gamma=10^{\text{-}3}$, but there is a reduction in the number of grid points. 
With this said, DDSG requires roughly $10$ times fewer grid points to attain equivalent Euler adaptive SG errors. 
In the bottom panel of Figure~\ref{fig:10}, we show the average and maximum errors for the non-smooth IRBC model using SG, adaptive SG, and the DDSG method at varying SG adaptivity coefficients. 
Unlike the previous case, the non-smooth IRBC model requires significantly higher SG refinement levels to represent its non-smooth policy function adequately. 
Two tests were conducted using the DDSG method with $\ekk=2$, one using $\ell=8$ with and $\epsilon_\gamma= 10^{\text{-}2}$, and the other with $\ell=10$ and $5 \times 10^{\text{-}3}$. 
Notice that in high-dimensions, at such high SG refinement levels, 
the number of grid points would almost surely render a model uncomputable for adaptive SGs, even if run on modern supercomputer facilities. 
For example, using an SG at $\ell=10$, a $20$-dimensional model would consist of about $5$ billion grid points. 
As observed in the right bottom panel, DDSG requires higher refinement levels to sufficiently decrease the maximum Euler error. 
Even at this refinement level, in comparison to adaptive SG at $\ell=8$ and $\epsilon_\gamma=10^{\text{-}2}$, the DDSG method allows for a significant reduction in the Euler error with half the number of grid points of adaptive SG. 

\subsection{Large-scale models}\label{sec:5.3}
Based on the analysis conducted in the previous section, we proceed here by adopting the DDSG parameters noted above as a baseline for solving a set of large-scale IRBC models with up to $300$ and $60$ dimensions for the smooth and non-smooth IRBC models, respectively.
At such dimensions, adaptive SGs would not be a fitting numerical approach due to the sheer number of grid points. 
The effect of this massive increase in the number of grid points is proportional to an increase in the computation time, which quickly surpasses a day of runtime on a high-performance computer. 
In Table~\ref{tab:2}, we show the results for the two IRBC model variants.
Firstly, we achieve low average and maximum Euler errors for the smooth model for all test cases.
For the $300$-dimensional model, we have used a lower SG refinement level of $3$, compared to $4$ for the $100$- and $200$-dimensional models. 
To compensate for this lower refinement level, we use $\epsilon_\gamma=10^{\text{-}6}$. 
With this said, we can see that even for the $300$-dimensional case, the proposed framework is sufficient to achieve $-2.89$ and $-1.78$ for the average and maximum Euler errors, respectively.
For comparative purposes, we show the number of SG grid points at refinement level $4$. 
There is roughly a four-orders-of-magnitude difference between the required number of grid points between SG and the DDSG. 
Using adaptive SGs will undoubtedly decrease the number of grid points. However, dimensions $>100$ remain uncomputable using adaptive SG. 
The per iteration runtimes of the smooth IRBC model tests are $0.5$, $1.6$ and $4.2$ hours using $100$, $200$ and $300$ nodes, for model dimensions $100$, $200$ and $300$, respectively.
For the non-smooth IRBC models, we look at model dimensions of $20$, $40$, and $60$.
Compared to the smooth model, the number of grid points required is significantly larger, as we need a much denser grid to capture the strong nonlinearities in the policy functions.
We can efficiently alleviate this problem by using DDSG with a refinement level of $10$. 
The shortfall of a pure SG becomes apparent when we look at a comparative number of SG grid points at the same refinement level, surpassing trillions of grid points in a $60$-dimensional model with level $10$. 
Using adaptive SGs can provide some degree of efficiency in lower-dimensions, for example,~\cite{brummscheidegger_2017} show that a $20$-dimensional model can be solved using roughly $10^4$ grid points. 
With this said, the DDSG approximation method required more than half the grid points to achieve the same error metrics. 
Furthermore, in a higher dimension, the number of grid points for both SG and adaptive SG will increase significantly, rendering the model uncomputable on contemporary supercomputers. 
The per-iteration-runtimes of the kink model are $1.8$, $9.9$, and $16.4$ hours using $38$, $156$, and $354$ nodes, for model dimensions $20$, $40$ and $60$, respectively. 

\begin{table}[t]%
\centering
\caption{Large--scale results for the smooth and non-smooth IRBC model.}
\scalebox{0.9}{
\begin{tabular}{c c| c c c c| c c| c c} 
\multicolumn{2}{c}{\textbf{Model}} &\multicolumn{4}{c}{\textbf{DDSG Parameters}} & \multicolumn{2}{c}{\textbf{Grid Points}}  & 
\multicolumn{2}{c}{\textbf{Euler Error}} \\ 
\multicolumn{1}{c}{Type}&\multicolumn{1}{c}{$d$} & \multicolumn{1}{c}{$\ekk$} & \multicolumn{1}{c}{$\epsilon_\eta=\epsilon_\rho$} &  \multicolumn{1}{c}{$\ell$} & \multicolumn{1}{c}{$\epsilon_\gamma$} & \multicolumn{1}{c}{DDSG} & \multicolumn{1}{c}{SG$^*$} & \multicolumn{1}{c}{Avg.} & \multicolumn{1}{c}{Max.} \\ \hline \rule{0pt}{2.6ex}\rule[-0.9ex]{0pt}{0pt} 
\multirow{3}{*}{smooth} &
100  & 1& $ 10^{\text{-}4}$& 4& $10^{\text{-}3}$ & 8.1$\times 10^2$& 1.4$\times 10^6$ & -3.35& -2.21 \\ 
&200  & 1& $ 10^{\text{-}4}$& 4& $10^{\text{-}3}$ & 1.6$\times 10^3$& 1.1$\times 10^7$ & -2.95& -2.15 \\
&300  & 1& $ 10^{\text{-}4}$& 3& $ 10^{\text{-}6}$        & 1.5$\times 10^3$& 3.6$\times 10^7$ & -2.89& -1.78 \\ \hline \rule{0pt}{2.6ex}\rule[-0.9ex]{0pt}{0pt} 
\multirow{3}{*}{non-smooth} &
20  & 2 & $ 10^{\text{-}4}$ & 10 & 5$\times 10^{\text{-}3}$ & 4.3$\times 10^{3}$ & 1.4$\times 10^{9}$ & -2.79 & -1.92 \\ 
&40  & 2 & $ 10^{\text{-}4}$ & 10 & 5$\times 10^{\text{-}3}$ & 1.7$\times 10^{4}$ & $\gg 10^{10}$       & -2.71 & -1.98 \\
&60  & 2 & $ 10^{\text{-}4}$ & 10 & 5$\times 10^{\text{-}3}$ & 3.1$\times 10^{4}$ & $\gg  10^{10}$      & -2.84 & -1.96 \\
\end{tabular}
}
\begin{tablenotes}
\item{\small{\textit{Average and maximum Euler errors for smooth IRBC models using the DDSG time iteration method.
Note that no SG tests could be conducted at such high dimensions.
$^*$The reported SG grid points are for comparison purposes and are reported using $\ell=4$ and $10$ for the smooth and non-smooth model, respectively. 
}
}}
\end{tablenotes}
\label{tab:2}
\end{table}
%

\section{Conclusions} \label{sec:6}
We introduced a computational framework to solve large-scale dynamic stochastic economic models on practical time scales. 
As a secondary benefit, it can serve as an a priori analysis tool to shed light on the model's complexity. 
At the core of the methodology is the DDSG function approximation that combines an HDMR technique with adaptive SGs.
We parallelized the DDSG method by leveraging the intrinsic separability in the computation, embedded it in a time iteration algorithm, and deployed it on the Cray XC$50$ system installed at CSCS. 
Our numerical experiments---that is, solving a set of smooth and non-smooth IRBC models, showed a speedup of $10$ times in comparison to state-of-the-art adaptive SGs in cases of mid-scale models, where both methods were applicable. 
In addition, we showed that even for a relatively small $50$-dimensional model, the proposed framework provides excellent strong scaling up $1,000$ compute nodes.
Furthermore, we demonstrated that we can compute global solutions to IRBC models with at least $300$ continuous dimensions in only a few hours.
This is a substantial improvement over the previous literature and opens new possibilities for a richer set of model specifications. 
Finally, note that the scope of the presented method is not restricted to models that are recursively formulated via first-order conditions, but more broadly to high-dimensional models that can be characterized in the functional equation~\eqref{eq:DynamicEq}. The latter also nests the common characterizations of recursive equilibria in discrete time, where $p$ is the value function, and the operator $\mathcal{H}$ captures the Bellman equation it has to satisfy---or the Hamilton-Jacobi-Bellman equation in continuous time~\cite{FERNANDEZVILLAVERDE2016527}. However, tackling such models with the proposed method is subject to further research.

\section*{Acknowledgments}

We thank Lorenzo Bretscher, Johannes Brumm, Felix K\"ubler, Philipp Renner, Olaf Schenk, Karl Schmedders, Tony Smith, Fabio Trojani, and seminar participants at the University of Geneva, the University of Lausanne, Università della Svizzera italiana, the Russian Presidential Academy of Science, Stanford University, the University of Zurich for their extremely valuable comments.

\bibliographystyle{siamplain}
\bibliography{bib_econ_nourl.bib} 
\end{document}

%% file: ex_shared.tex
\usepackage{lipsum}
\usepackage{amsfonts}
\usepackage{graphicx}
\usepackage{epstopdf}
\usepackage{algorithmic}
\ifpdf
  \DeclareGraphicsExtensions{.eps,.pdf,.png,.jpg}
\else
  \DeclareGraphicsExtensions{.eps}
\fi

\newsiamremark{remark}{Remark}
\newsiamremark{hypothesis}{Hypothesis}
\crefname{hypothesis}{Hypothesis}{Hypotheses}
\newsiamthm{claim}{Claim}

\headers{High-Dimensional Dynamic Stochastic Model Representation}{Aryan Eftekhari, Simon Scheidegger}

\title{High-Dimensional Dynamic Stochastic Model Representation
\thanks{Submitted to the editors on January 6, 2021.
\funding{This work was generously supported by grants from the Swiss National Supercomputing Centre (CSCS) under project IDs s885 and s995, the Swiss Platform for Advanced Scientific Computing (PASC) under project ID ``Computing equilibria in heterogeneous agent macro models on contemporary HPC platforms'', and the Swiss National Science Foundation under project IDs ``Can Economic Policy Mitigate Climate-Change?'' and ``New methods for asset pricing with frictions''.}
}}

\author{Aryan Eftekhari
  \thanks{Institute of Computational Science, Università della Svizzera italiana, Lugano, Switzerland, and Department of Economics, University of Lausanne, 
        	Switzerland
  (\email{aryan.eftekhari@usi.ch}).}
\and Simon Scheidegger 
\thanks{Department of Economics, University of Lausanne,
        	Switzerland, and Enterprise for Society (E4S) 
  (\email{simon.scheidegger@unil.ch}).}
  }

\usepackage{amsopn}


%% file: ex_article.bbl
\begin{thebibliography}{10}

\bibitem{azinovic_et_al_2019}
{\sc M.~Azinovic, L.~Gaegauf, and S.~Scheidegger}, {\em Deep equilibrium nets},
   (2019), \url{http://dx.doi.org/10.2139/ssrn.3393482}.

\bibitem{Bellman1961AdaptiveTour}
{\sc R.~E. Bellman}, {\em Adaptive Control Processes: A Guided Tour}, Princeton
  University Press, 1961.

\bibitem{bengui2013}
{\sc J.~Bengui, E.~G. Mendoza, and V.~Quadrini}, {\em {Capital mobility and
  international sharing of cyclical risk}}, Journal of Monetary Economics, 60
  (2013), pp.~42--62.

\bibitem{SG_in_econ_handbook}
{\sc J.~Brumm, C.~Krause, A.~Schaab, and S.~Scheidegger}, {\em Sparse grids for
  dynamic economic models},  (2021), \url{https://ssrn.com/abstract=3979412}.

\bibitem{Brumm201512}
{\sc J.~Brumm, D.~Mikushin, S.~Scheidegger, and O.~Schenk}, {\em Scalable
  high-dimensional dynamic stochastic economic modeling}, Journal of
  Computational Science, 11 (2015), pp.~12 -- 25.

\bibitem{brummscheidegger_2017}
{\sc J.~Brumm and S.~Scheidegger}, {\em Using adaptive sparse grids to solve
  high-dimensional dynamic models}, Econometrica, 85 (2017), pp.~1575--1612.

\bibitem{Bungartz2003MultivariateGrids}
{\sc H.~J. Bungartz and S.~Dirnstorfer}, {\em {Multivariate Quadrature on
  Adaptive Sparse Grids}}, Computing (Vienna/New York), 71 (2003), pp.~89--114.

\bibitem{Bungartz.Griebel:2004}
{\sc H.-J. Bungartz and M.~Griebel}, {\em Sparse grids}, Acta Numerica, 13
  (2004), pp.~1--123.

\bibitem{coleman1990}
{\sc W.~J. Coleman}, {\em Solving the stochastic growth model by
  policy-function iteration}, Journal of Business \& Economic Statistics, 8
  (1990), pp.~27--29.

\bibitem{RePEc:eee:dyncon:v:35:y:2011:i:2:p:175-177}
{\sc W.~J. Den~Haan, K.~L. Judd, and M.~Juillard}, {\em Computational suite of
  models with heterogeneous agents ii: Multi-country real business cycle
  models}, Journal of Economic Dynamics and Control, 35 (2011), pp.~175--177.

\bibitem{den1990solving}
{\sc W.~J. Den~Haan and A.~Marcet}, {\em Solving the stochastic growth model by
  parameterizing expectations}, Journal of Business \& Economic Statistics, 8
  (1990), pp.~31--34.

\bibitem{Eftekhari:2017:PDD:3093172.3093234}
{\sc A.~Eftekhari, S.~Scheidegger, and O.~Schenk}, {\em Parallelized
  dimensional decomposition for large-scale dynamic stochastic economic
  models}, in Proceedings of the Platform for Advanced Scientific Computing
  Conference, PASC '17, New York, NY, USA, 2017, ACM, pp.~9:1--9:11.

\bibitem{fernandez-villaverdehurtadonuno_2019_WP}
{\sc J.~Fernández-Villaverde, S.~Hurtado, and G.~Nuno}, {\em Financial
  frictions and the wealth distribution}.
\newblock Working paper, 2019.

\bibitem{FERNANDEZVILLAVERDE2016527}
{\sc J.~Fernández-Villaverde, J.~Rubio-Ramírez, and F.~Schorfheide}, {\em
  Chapter 9 - solution and estimation methods for dsge models}, vol.~2 of
  Handbook of Macroeconomics, Elsevier, 2016, pp.~527--724.

\bibitem{GAO20103274}
{\sc Z.~Gao and J.~S. Hesthaven}, {\em On anova expansions and strategies for
  choosing the anchor point}, Applied Mathematics and Computation, 217 (2010),
  pp.~3274--3285.

\bibitem{Holtz2010SparseInsurance}
{\sc M.~Holtz}, {\em {Sparse Grid Quadrature in High Dimensions with
  Applications in Finance and Insurance}}, Lecture Notes in Computational
  Science and Engineering, 77 (2010), pp.~1--192.

\bibitem{Hooker}
{\sc G.~Hooker}, {\em Generalized functional anova diagnostics for
  high-dimensional functions of dependent variables}, Journal of Computational
  and Graphical Statistics, 16 (2007), pp.~709--732.

\bibitem{Hooker2007GeneralizedVariables}
{\sc G.~Hooker}, {\em {Generalized Functional ANOVA Diagnostics for
  High-Dimensional Functions of Dependent Variables}}, Journal of Computational
  and Graphical Statistics, 16 (2007), pp.~709--732.

\bibitem{Hsiang2020}
{\sc S.~Hsiang, D.~Allen, S.~Annan-Phan, K.~Bell, I.~Bolliger, T.~C.~S. Chong,
  H.~Druckenmiller, L.~Huang, A.~Hultgren, E.~Krasovich, P.~Lau, J.~Lee,
  E.~Rolf, J.~Tseng, and T.~Wu}, {\em The effect of large-scale anti-contagion
  policies on the covid-19 pandemic}, Nature, 585 (2020).

\bibitem{judd1998numerical}
{\sc K.~Judd}, {\em Numerical Methods in Economics}, Scientific and
  Engineering, MIT Press, 1998.

\bibitem{Judd1998NumericalEconomicsb}
{\sc K.~L. Judd}, {\em {Numerical methods in economics}}, vol.~1, The MIT
  press, 1998.

\bibitem{Juillard2011178}
{\sc M.~Juillard and S.~Villemot}, {\em Multi-country real business cycle
  models: Accuracy tests and test bench}, Journal of Economic Dynamics and
  Control, 35 (2011), pp.~178 -- 185.

\bibitem{kaplanetal_2018}
{\sc G.~Kaplan, B.~Moll, and G.~L. Violante}, {\em Monetary policy according to
  hank}, American Economic Review, 108 (2018), pp.~697--743.

\bibitem{Kollmann2011186}
{\sc R.~Kollmann, S.~Maliar, B.~A. Malin, and P.~Pichler}, {\em {Comparison of
  solutions to the multi-country Real Business Cycle model}}, Journal of
  Economic Dynamics and Control, 35 (2011), pp.~186--202.

\bibitem{kotlikoffetal_2020_WP}
{\sc L.~Kotlikoff, F.~Kubler, A.~Polbin, and S.~Scheidegger}, {\em
  {Pareto-improving carbon-risk taxation}}, Economic Policy, 36 (2021),
  pp.~551--589.

\bibitem{Krueger20041411}
{\sc D.~Krueger and F.~Kubler}, {\em Computing equilibrium in {OLG} models with
  stochastic production}, Journal of Economic Dynamics and Control, 28 (2004),
  pp.~1411 -- 1436.

\bibitem{kruegeretal_2016}
{\sc D.~Krueger, K.~Mitman, and F.~Perri}, {\em Chapter 11 - macroeconomics and
  household heterogeneity}, vol.~2 of Handbook of Macroeconomics, Elsevier,
  2016, pp.~843 -- 921.

\bibitem{kublerscheidegger_2018_WP}
{\sc F.~Kubler and S.~Scheidegger}, {\em Self-justified equilibria: Existence
  and computation},  (2019), \url{https://ssrn.com/abstract=3494876}.

\bibitem{kublerschmedders_2003}
{\sc F.~Kubler and K.~Schmedders}, {\em Stationary equilibria in asset-pricing
  models with incomplete markets and collateral}, Econometrica, 71 (2003),
  pp.~1767--1793.

\bibitem{Kuo2009OnFunctions}
{\sc F.~Y. Kuo, I.~H. Sloan, G.~W. Wasilkowski, and H.~Wo{\'{z}}niakowski},
  {\em {On decompositions of multivariate functions}}, Mathematics of
  Computation, 79 (2009), pp.~953--966.

\bibitem{Li2012GeneralVariables}
{\sc G.~Li and H.~Rabitz}, {\em {General formulation of HDMR component
  functions with independent and correlated variables}}, Journal of
  Mathematical Chemistry, 50 (2012), pp.~99--130.

\bibitem{genyuan}
{\sc G.~Li, C.~Rosenthal, and H.~Rabitz}, {\em High dimensional model
  representations}, The Journal of Physical Chemistry A, 105 (2001),
  pp.~7765--7777.

\bibitem{ljungqvist2004recursive}
{\sc L.~Ljungqvist and T.~J. Sargent}, {\em Recursive macroeconomic theory},
  Mit Press, 2004.

\bibitem{Ma:2009:AHS:1514432.1514547}
{\sc X.~Ma and N.~Zabaras}, {\em An adaptive hierarchical sparse grid
  collocation algorithm for the solution of stochastic differential equations},
  J. Comput. Phys., 228 (2009), pp.~3084--3113.

\bibitem{Ma2010AnEquations}
{\sc X.~Ma and N.~Zabaras}, {\em {An adaptive high-dimensional stochastic model
  representation technique for the solution of stochastic partial differential
  equations}}, Journal of Computational Physics, 229 (2010), pp.~3884--3915.

\bibitem{MALIAR2014325}
{\sc L.~Maliar and S.~Maliar}, {\em Chapter 7 - numerical methods for
  large-scale dynamic economic models}, in Handbook of Computational Economics
  Vol. 3, K.~Schmedders and K.~L. Judd, eds., vol.~3 of Handbook of
  Computational Economics, Elsevier, 2014, pp.~325 -- 477.

\bibitem{MALIAR202176}
{\sc L.~Maliar, S.~Maliar, and P.~Winant}, {\em Deep learning for solving
  dynamic economic models.}, Journal of Monetary Economics, 122 (2021),
  pp.~76--101.

\bibitem{memoization}
{\sc D.~MICHIE}, {\em ``memo''functions and machine learning}, Nature, 218
  (1968), pp.~19--22.

\bibitem{DBLP:journals/corr/abs-1906-02005}
{\sc V.~Minh Nguyen-Thanh, L.~Trong Khiem~Nguyen, T.~Rabczuk, and X.~Zhuang},
  {\em A surrogate model for computational homogenization of elastostatics at
  finite strain using high-dimensional model representation-based neural
  network}, International Journal for Numerical Methods in Engineering,
  (2020).

\bibitem{ppopp027s-murarasu}
{\sc A.~Murarasu, J.~Weidendorfer, G.~Buse, D.~Butnaru, and D.~Pfl\"{u}eger},
  {\em Compact data structure and parallel alogrithms for the sparse grid
  technique}, in 16th ACM SIGPLAN Symposium on Principles and Practice of
  Parallel Programming, 2011.

\bibitem{pflueger10spatially}
{\sc D.~Pfl{\"{u}}ger}, {\em Spatially Adaptive Sparse Grids for
  High-Dimensional Problems}, PhD thesis, M{\"{u}}nchen, Aug. 2010.

\bibitem{Pfluger2010SpatiallyProblems}
{\sc D.~Pflüger, B.~Peherstorfer, and H.-J. Bungartz}, {\em Spatially adaptive
  sparse grids for high-dimensional data-driven problems}, Journal of
  Complexity, 26 (2010), pp.~508 -- 522.

\bibitem{RABITZ199911}
{\sc H.~Rabitz, {\"{O}}.~F. Ali{\c{s}}, J.~Shorter, and K.~Shim}, {\em
  Efficient input--output model representations}, Computer Physics
  Communications, 117 (1999), pp.~11--20.

\bibitem{Rabitz1999GeneralRepresentations}
{\sc H.~Rabitz and {\"{O}}.~F. Aliş}, {\em {General foundations of
  high‐dimensional model representations}}, Journal of Mathematical
  Chemistry, 25 (1999), pp.~197--233.

\bibitem{rennerscheidegger_2020}
{\sc P.~Renner and S.~Scheidegger}, {\em Machine learning for dynamic incentive
  problems},  (2018).

\bibitem{scheideggerbilinois_2019}
{\sc S.~Scheidegger and I.~Bilionis}, {\em Machine learning for
  high-dimensional dynamic stochastic economies}, Journal of Computational
  Science, 33 (2019), pp.~68 -- 82.

\bibitem{IPDPS2018}
{\sc S.~Scheidegger, D.~Mikushin, F.~Kubler, and O.~Schenk}, {\em Rethinking
  large-scale economic modeling for efficiency: Optimizations for gpu and xeon
  phi clusters}, in 2018 IEEE International Parallel and Distributed Processing
  Symposium (IPDPS), May 2018, pp.~610--619.

\bibitem{scheidegger.treccani.20138}
{\sc S.~Scheidegger and A.~Treccani}, {\em Pricing american options under
  high-dimensional models with recursive adaptive sparse expectations}, Journal
  of Financial Econometrics,  (2018).

\bibitem{Sobol2003TheoremsRepresentation}
{\sc I.~M. Sobol}, {\em {Theorems and examples on high dimensional model
  representation}}, Reliability Engineering {\&} System Safety, 79 (2003),
  pp.~187--193.

\bibitem{stokey1989recursive}
{\sc N.~L. Stokey, R.~E. Lucas, and E.~C. Prescott}, {\em Recursive Methods in
  Economic Dynamics}, Harvard University Press, 1989.

\bibitem{Wachter2005OnProgramming}
{\sc A.~W\"{a}chter and L.~T. Biegler}, {\em {On the implementation of an
  interior-point filter line-search algorithm for large-scale nonlinear
  programming}}, Math. Program., Ser. A,  (2005).

\bibitem{Wang2008OnRepresentation}
{\sc X.~Wang}, {\em {On the approximation error in high dimensional model
  representation}}, Proceedings of the 2008 Winter Simulation Conference,
  (2008), pp.~453--462.

\bibitem{Yang2012AdaptiveFlows}
{\sc X.~Yang, M.~Choi, G.~Lin, and G.~E. Karniadakis}, {\em {Adaptive ANOVA
  decomposition of stochastic incompressible and compressible flows}}, Journal
  of Computational Physics, 231 (2012), pp.~1587--1614.

\end{thebibliography}
